\documentclass[pra,twocolumn, superscriptaddress]{revtex4-2}
\usepackage{graphicx}
\usepackage{epsfig}
\usepackage{amsfonts}
\usepackage{amsmath}
\usepackage{amssymb}
\usepackage{dsfont}
\usepackage{amsthm}
\usepackage{color}
\usepackage{mathrsfs}
\usepackage{enumitem}
\usepackage[colorlinks=true,linkcolor=blue,citecolor=blue,urlcolor=blue]{hyperref}
\usepackage{comment}
\usepackage{braket}
\usepackage{physics}
\usepackage{xr}
\usepackage[dvipsnames]{xcolor}
\usepackage{soul}
\definecolor{dgreen}{rgb}{0.0, 0.6, 0.0}
\newcommand{\argp}[1]{\left( #1 \right)}
\newcommand{\args}[1]{\left[ #1 \right]}

\newcommand{\add}[1]{\textcolor{dgreen}{#1}}

\begin{document}

\title{Long-distance entanglement sharing using hybrid states of discrete and continuous variables}

\author{Soumyakanti Bose}
\email{soumyakanti.bose09@gmail.com}
\affiliation{Department of Physics \& Astronomy, Seoul National University, Gwanak-ro 1, Gwanak-gu, Seoul 08826, Korea}
\author{Jaskaran Singh}
\email{jaskaran@us.es}
\affiliation{Departamento de F\'{\i}sica Aplicada II, Universidad de Sevilla, E-41012 Sevilla, Spain}
\affiliation{Department of Physics and Center for Quantum Frontiers of Research \&
Technology (QFort), National Cheng Kung University, Tainan 701, Taiwan}
\author{Ad\'an Cabello}
\email{adan@us.es}
\affiliation{Departamento de F\'{\i}sica Aplicada II, Universidad de Sevilla, E-41012 Sevilla, Spain}
\affiliation{Instituto Carlos~I de F\'{\i}sica Te\'orica y Computacional, Universidad de Sevilla, E-41012 Sevilla, Spain}
\author{Hyunseok Jeong}
\email{h.jeong37@gmail.com}
\affiliation{Department of Physics \& Astronomy, Seoul National University, Gwanak-ro 1, Gwanak-gu, Seoul 08826, Korea}


\begin{abstract}
We introduce a feasible scheme to produce high-rate long-distance entanglement which uses hybrid entanglement (HE) between continuous variables (CV) and discrete variables (DV). 
We show that HE can effectively remove the experimental limitations of existing CV and DV systems to produce long range entanglement. 
We benchmark the resulting DV entangled states using an entanglement-based quantum key distribution (EB-QKD) protocol.
We show that, using HE states, EB-QKD is possible with standard telecommunication fibers for $300$ km.
The key idea is using the CV part, which can be adjusted to be robust against photon losses, for increasing the transmission distance, while using the DV part for achieving high secure key rates. 
Our results point out that HE states provide a clear advantage for practical long-distance and high-rate entanglement generation that may lead to further applications in quantum information processing.
\end{abstract}


\maketitle



\section{Introduction}
Generation of high-rate entanglement between distant locations is crucial for fundamental tests of quantum theory and many applications. For example, it is needed for extending the current distances and rates of
loophole-free Bell tests~\cite{HBD15,RBG17}, 
quantum steering~\cite{BSF12}, and quantum teleportation~\cite{MHS12}, which so far are only feasible for relatively short ranges. 
It is also needed for increasing the transmission distance and the key rate of entanglement-based quantum key distribution (EB-QKD) protocols, most notably device-independent QKD~\cite{NDN22,ZVR22,LZZ22}, which currently suffers from both these issues. 
Moreover, higher-rates in distant locations will also allow us to achieve higher detection efficiencies (which are needed both for loophole-free Bell tests and device-independent QKD) by means of heralded qubit amplifiers \cite{Gisin2010PRL} or photonic precertification schemes \cite{Cabello2012PRX,Meyer-Scott2016PRL,Leger2022arXiv}, whose practicality is currently limited by the rates achieved after transmission. 

A benchmark of high-rate entanglement over long distances, from an operational perspective, can be set by its performance in an information processing task such as an EB-QKD protocol.
These protocols can be broadly classified into two distinct classes: (i) those using discrete variable (DV) entangled states and (ii) those that use continuous variable (CV) entangled states, where each class has its own set of advantages and limitations~\cite{DLQ16, XCQ15, POS15_reply}.
As an example, DV EB-QKD protocols offer composable security proofs with good key rate, but they require precise Bell-state or single-photon measurements at extremely low temperatures, which are hard to perform even in laboratory conditions.
On the other hand, CV EB-QKD protocols generally require Gaussian states which are comparatively easier to prepare, but their performance is limited by the requirement of almost ideal homodyne detectors at telecommunication wavelength \cite{KSB19, SB21, XCQ15}.
As a consequence, despite an extensive theoretical and experimental analysis on both types of systems, the quest for an optimal physical system which can be potentially used to share high-rate entanglement remains open.

Nonetheless, there exists a different class of physical systems where the entanglement is between CV and DV systems and are formally known as hybrid entangled (HE) states~\cite{J05, LJZ06, LGZ06, HLS11, HRS12, LAA21}. 
These strongly correlated~\cite{CHZ02, PLJ12, KJ13} cross-system entangled states play a crucial role in various quantum information processing tasks, including quantum computation, communication, and tests of Bell non-locality~\cite{LJ13, ANV15, OTJ20, OTL21, BJ22, HM22, KJ13, Sheng2013, YLim2016}, and have been efficiently generated in a wide range of experimental setups~\cite{JZK14, MHL14, USP17, SUT18, DAG23}. 
Consequently, it becomes interesting to observe whether such hybrid states can be used to share entanglement among distant locations without the limitations faced by CV and DV systems.

\begin{figure*}
    \centering
    \includegraphics[scale=0.8]{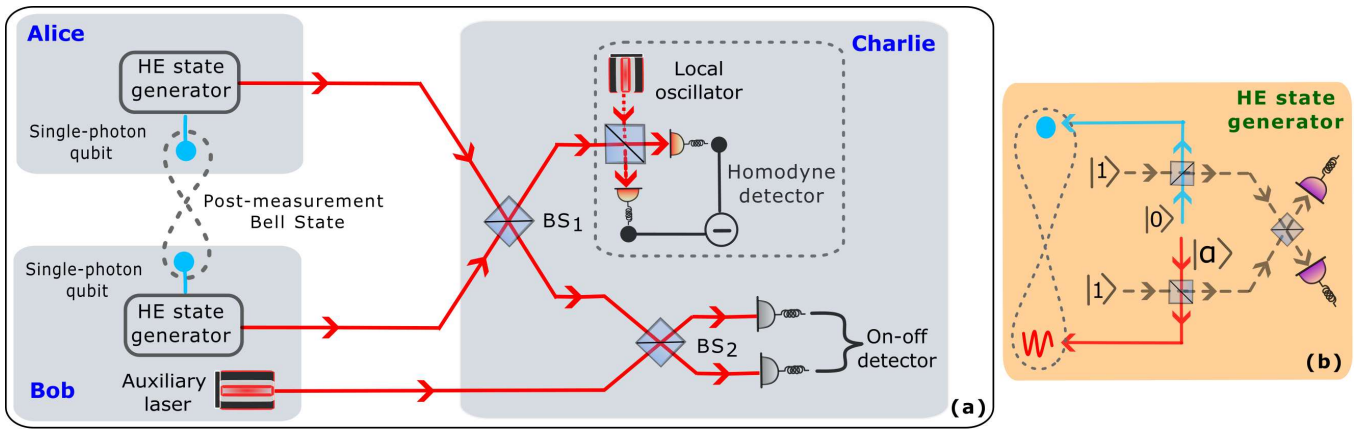}
    \caption{(a) Scheme for generating DV entangled states between Alice and Bob using HE states. 
    The DV part (cyan) and the CV part (red) of the HE state stand for the single-photon state and coherent state, respectively. 
    Alice and Bob send the CV part of their individual HE states to Charlie, who then mixes the incoming signals at a balanced beam splitter (BS$_1$), and uses one of the output modes for homodyne measurement with efficiency $\eta_h$.
    The other outgoing signal of BS$_1$ is used for a post-selection measurement by on-off detectors with efficiency $\eta_0$, after mixing it at another balanced BS (BS$_2$) with the additional coherent signal sent by Bob. 
    Upon declaration of the results by Charlie, Alice and Bob obtain a DV entangled pair which is used for secure key generation.
    (b) Scheme for generating HE states.
    Two ancilla single photons (gray, dashed line) are mixed with vacuum and coherent states at the two BSs.
    The outgoing ancilla photons are then mixed with each other at a second BS.
    When the detector placed at the output of the second BS clicks, then the HE state between single-photon and the coherent state is obtained.}
    \label{fig:protocol}
\end{figure*} 

Here, we propose a scheme based on HE states as an initial resource which produces high-rate DV entanglement between extremely far apart locations. 
We provide a characterization of such states and show that it is possible to share entanglement between locations which are hundreds of kms apart.
We further assess the quality of shared entanglement in the context of EB-QKD.
We show that, by bringing forth the best of both CV and DV systems, with HE states, it is possible to achieve secure key rate at a distance of $300$ km by using practical homodyne detectors with efficiency $\eta_h = 0.55$~\cite{QLP15, JKL13, ZMC22, ZHC19, ZCP20} (which is a reasonable value at telecommunication wavelengths~\mbox{\cite{JKL13}}) and on-off detectors with efficiency $\eta_0 = 0.8$~\cite{onoff_detect}.  
Note that we use the key rates and transmission distances only to quantify the quality of the entanglement; our central goal is to show the advantage of using HE states to achieve entanglement over longer distances, which is a crucial tool enabling a wide range of fundamental tests in physics and quantum information processing applications.

Our scheme hinges on generating a single-photon DV entangled state between two distant parties by exploiting CV entanglement swapping \cite{YLim2016} by a third party located midway.
It offers three major advantages as compared to earlier CV and DV EB-QKD protocols.
These are: (i) Elimination of major limiting factors of DV EB-QKD, which include high precision Bell state or single-photon measurements as well as the photon-number-splitting attack by an eavesdropper by considering entanglement swapping over the CV system.
(ii) Elimination of the requirement of near-unit efficiency for the homodyne detectors used for key generation in CV EB-QKD.
(iii) Long transmission distance at telecommunication wavelength stemming from the robustness of the multiphoton coherent state against transmission losses and using practical devices.

This article is organized as follows. In Sec.~\ref{sec:ent_sharing}, we provide some brief introduction to HE states. 
We then propose a protocol to share DV entanglement among distant locations by using HE states as an initial resource. 
We also characterize the resulting entanglement using logarithmic negativity and show that it can be non-zero even when the parties are separated hundreds of kilometers apart. 
In Sec.~\ref{sec:qkd}, we benchmark the usefulness of the resultant entangled states by demonstrating our scheme as an EB-QKD protocol using practical devices. 
In Sec.~\ref{sec:conc}, we conclude our results by arguing that our protocol provides a practical solution to the problem long distance entanglement generation which is a central requirement in several information processing tasks.

\section{Entanglement sharing with hybrid states}
\label{sec:ent_sharing}

In this section we first provide a brief description of HE states. Subsequently, we detail our protocol to share long distance entanglement using these states.

\subsection{Hybrid entangled states}
Let $\ket{0}$ and $\ket{1}$ correspond to photon number states in the Fock basis and $\ket{\alpha}$ correspond to a coherent state of a quantized light with coherent amplitude $\alpha$. 
For the remainder of this paper we will represent the number of photons as a DV system, while the coherent state represents a CV system. 
We define a HE state as an entangled pair, where the entanglement is between the DV and CV degrees of freedom.
Mathematically, such HE states can be written as
\begin{equation}
    \ket{\psi}_{a_1 a_2}=\frac{1}{\sqrt{2}}\left( \ket{0}_{a_1}\ket{\alpha}_{a_2} + \ket{1}_{a_1}\ket{-\alpha}_{a_2} \right)
    \label{eq:he_state},
\end{equation}
where $a_1$ and $a_2$ are the two modes pertaining to the DV and CV parts, respectively.

We stress that HE states with small coherent amplitudes ($\alpha\lesssim 1$) are experimentally available. 
They have been generated experimentally in various settings such as conditional photon subtraction on a coherent state~\cite{JZK14} as well as photon subtraction on two squeezed states~\cite{MHL14}
(see Appendix~\ref{append_sec:hesgeneration} for further details).
While these techniques produce HE states with non-unit probability, it should be noted that typical methods to generate standard entangled photon pairs, {\it e.g.}, the parametric down conversion, also does so non-deterministically.
In Fig. \ref{fig:protocol}(b), we outline the linear optics based schematic for generating HE-states as originally described in \cite{JZK14}.

\subsection{Protocol for entanglement sharing}
\label{sec:entanglement_sharing}
We consider two distant parties, Alice and Bob, each of them having access to bipartite HE states $\ket{\psi}_{a_1a_2}$ and $\ket{\psi}_{b_1b_2}$ given by Eq.~\eqref{eq:he_state}. 
We consider these as initial resource states which will be used to share a DV entangled state between the parties. 
We provide a step-by-step description of the protocol, schematically represented in Fig.~\ref{fig:protocol}(a), while a detailed mathematical calculation can be found in Appendix~\ref{append_sec:final state}.

{\bf Step 1:} Alice and Bob generate HE states $\ket{\psi}_{a_1a_2}$ and $\ket{\psi}_{b_1b_2}$ in their respective laboratories. 
Both parties transmit the CV part of their systems, corresponding to modes $a_2$ and $b_2$, respectively, to a third untrusted party, Charlie, who lies midway between them, through a lossy quantum channel with transmittance $T$ ($0\leq T\leq 1$). 
Additionally, Bob also transmits the state $\ket{\sqrt{2}\alpha}$ to Charlie separately through a similar quantum channel.
After passing through channels with transmission losses, Charlie receives the mode $a_2$ from Alice, the mode $b_2$ from Bob, and the additional state $\ket{\sqrt{2T}\alpha}$ from Bob, which we label by mode $c$. 
While a general quantum channel between the parties will comprise of both transmission loss and thermal noise, here, for simplicity, we only consider lossy quantum channels with no noise. 
In Appendix~\ref{subsec:hes_fid_charlie_channel} we demonstrate that the scenario involving a practical level of thermal noise closely matches our current findings.

\begin{figure}
    \centering
    \includegraphics[scale = 0.35,angle=-90]{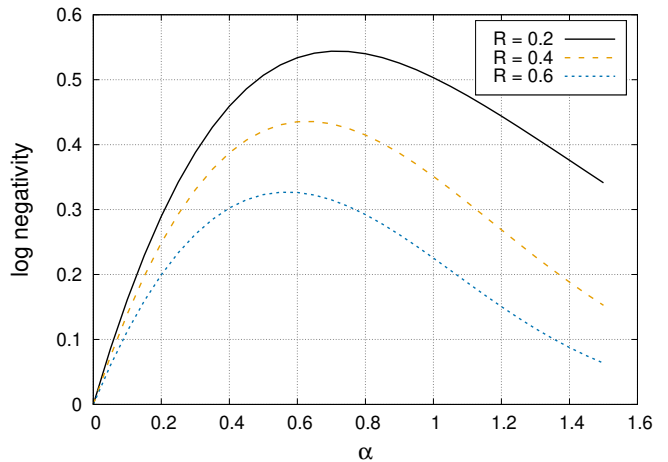}
    \caption{Logarithmic negativity of the HE state undergoing photon-loss over the CV part as a function of the coherent amplitude $\alpha$.
    $R$ ($0\leq R\leq 1$) stands for the normalized strength of loss.}
    \label{fig:logneg}
\end{figure}

The effect of a quantum state passing through a noisy channel can be seen as the system undergoing photon loss. 
In Fig.~\ref{fig:logneg}, we plot the logarithmic negativity of the HE state when its CV part undergoes photon loss as a function of the coherent amplitude $\alpha$ (see appendix \ref{append_sec:logneg_hybrid_photonloss} for detail). 
We find that there exists an optimal value of $\alpha$ for a fixed value of photon loss. We denote the photon loss fraction by $R$ such that $R=0$ and $R=1$ correspond to no photon loss and complete photon loss, respectively. 
For a significantly lossy channel, we find that the optimal value of $\alpha$ approaches $\alpha = 0.5$. This value becomes important when we benchmark the resultant DV entangled state by a EB-QKD protocol.

This behaviour of HE states can be qualitatively understood in terms of the interplay between entanglement and the fragility of the initial HE state.
Starting from the initial separable state at $\alpha=0$, the HE state becomes more entangled as $\alpha$ increases. 
An increase in $\alpha$ also corresponds to an increase in the average number of photons, which can be understood as an increase in the mean energy of the system.
However, with an increase in the mean energy, the state becomes more vulnerable to decoherence. 
This behaviour is similar to what is also shown in Ref.~\cite{LJ11} for superposition of coherent states, and the advantage of using small amplitudes under photon losses was demonstrated in the context of teleportation~\cite{Jonas2013}.
As a consequence, with increase in $\alpha$ beyond an optimal value, the HE state becomes extremely fragile under noise leading to a drop in entanglement when the multiphoton part passes through a noisy quantum channel. 

{\bf Step 2:} Next, Charlie mixes the two incoming modes $a_2$ and $b_2$ via a beam splitter (BS), labelled as BS$_1$ in Fig.~\ref{fig:protocol} with two output modes which we can label as $a_2^{'}$ and $b_2^{'}$.
In our protocol we are specifically interested in the vacuum state contributions from the mode $a_2^{'}$. 
To extract this contribution, Charlie mixes this mode though a second BS (BS$_2$) with mode $c$ with output modes labelled as $a_2^{''}$ and $c^{'}$.
Charlie now performs a projective measurement, $\mathcal{M} = \lbrace \Pi_0, \mathds{1}- \Pi_0\rbrace$, where $\Pi_0=(\mathds{1}-\ket{0}\bra{0})_{a_2^{''}}\otimes (\mathds{1}-\ket{0}\bra{0})_{c^{'}}$. 
This measurement is accomplished by using on-off detectors (that only detect the presence of photons) on each of the modes $a_2^{''}$ and $c^{'}$. 
Charlie then publicly announces the outcome of the projective measurement which is considered to be successful only if the result $\Pi_0$ is obtained, i.e., both detectors click. 
In that case, the protocol continues.
Otherwise, the measurement is deemed unsuccessful and the parties must repeat the aforementioned steps again. 
In order to model realistic detectors, we consider imperfect on-off detectors with efficiency $\eta_0$.

{\bf Step 3:} After a successful projective measurement (as dictated in Step 2), Charlie performs a homodyne measurement on  mode $b_2^{'}$ and, again, announces the results publicly.
We consider that homodyne measurements have efficiency $\eta_h$.

{\bf Step 4:} After a public announcement of the results of a successful projective measurement and the homodyne measurement by Charlie, Alice and Bob end up with the final normalized single-photon-Bell-state in modes $a_1$ and $b_1$ as
\begin{equation}
    \rho_{a_1b_1} = \frac{1}{2} \left[ 
    \begin{aligned}
        &\ket{01} \bra{01}+ \ket{10} \bra{10}\\
        &\quad+ h\left(g\ket{01}\bra{10} + g^*\ket{10}\bra{01}\right)
    \end{aligned}
    \right]
    \label{eq:final_state},
\end{equation}
with probability 
\begin{equation}
    P_0 = \frac{\left( 1 - e^{-\eta_o T\alpha^2} \right)^2}{2},
\end{equation}
where $h = e^{-4(1-T\eta_h)\alpha^2}$, $g = e^{4\mathsf{i}\sqrt{T\eta_h}\alpha p}$, $g^*$ is the conjugate of $g$, and $p$ is the result of the homodyne measurement.

\subsection{Shared DV-entanglement between the parties}
\label{subsec:entanglement_dv}
The entanglement of the final DV entangled state shared between the parties depends on a number of parameters. 
However, the quantities of most interest are the transmission length and the coherent amplitude. 
\begin{figure}[h]
    \centering
    \includegraphics[scale = 0.35, angle = -90]{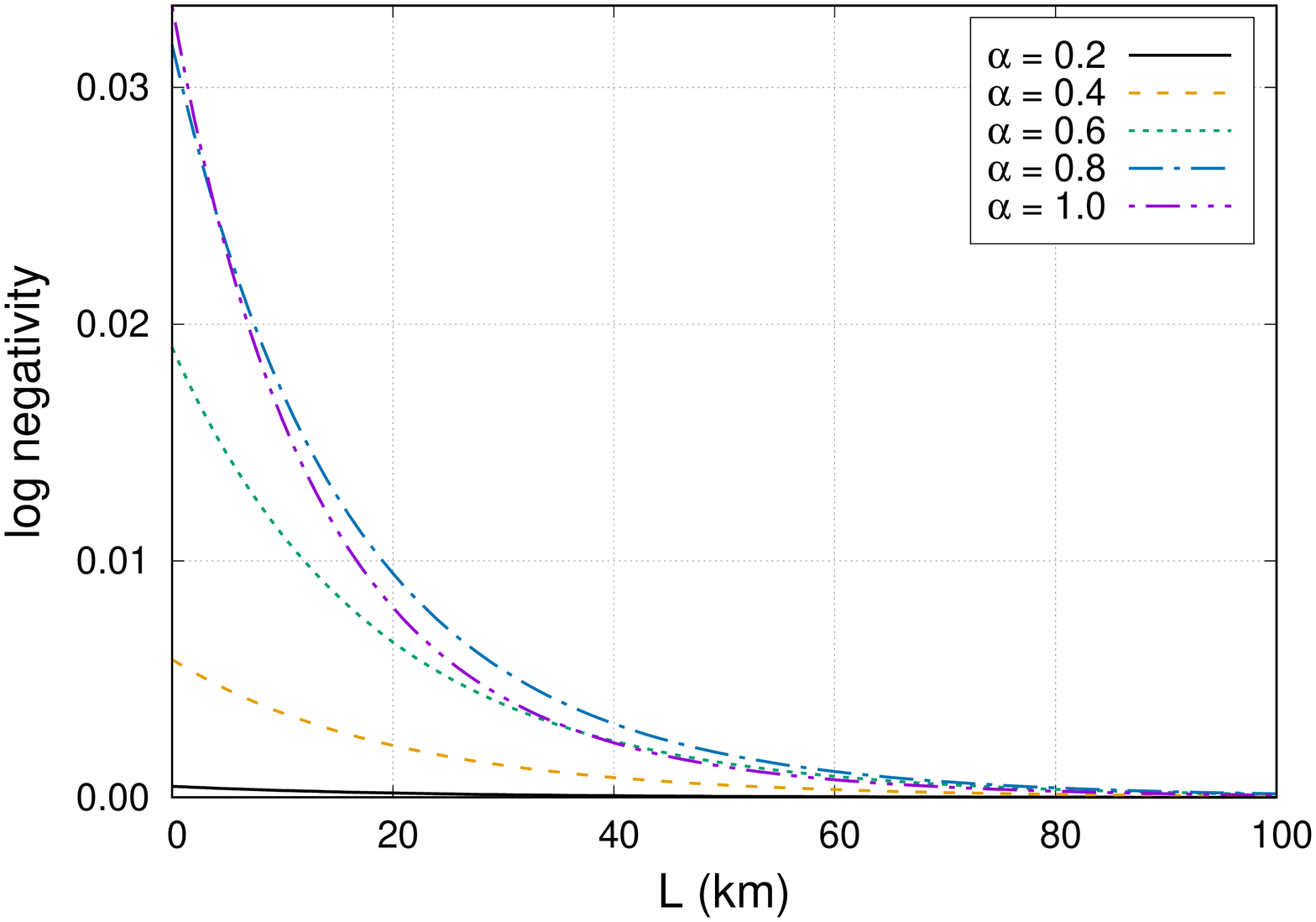}
    \caption{Logarithmic negativity of the state $\rho_{a_1b_1}$ as a function of the transmission distance ($L$) for different values of coherent amplitude $\alpha$. 
    We assume detection efficiencies $\eta_h = 0.55$ for the homodyne detectors, $\eta_0 = 0.8$ for the on-off detectors, and $p = \frac{\pi}{2}$.}
    \label{fig:logneg_length_final_state}
\end{figure}

\begin{figure}[h]
    \centering
    \includegraphics[scale = 0.35, angle = -90]{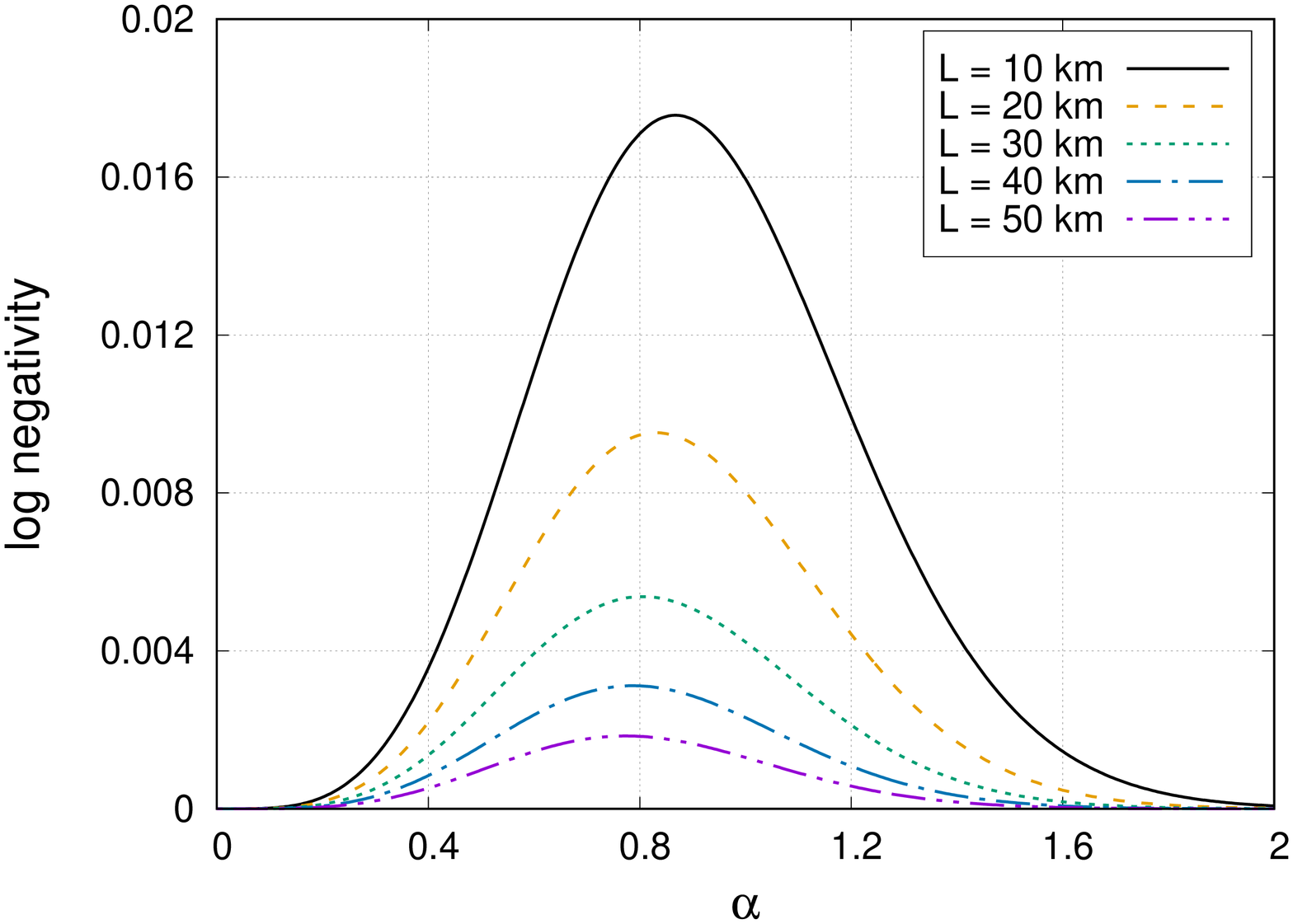}
    \caption{Logarithmic negativity of the state $\rho_{a_1b_1}$ as a function of the transmission distance $L$ for different values of coherent amplitude $\alpha$. 
    We assume detection efficiencies $\eta_h = 0.55$ for the homodyne detectors, $\eta_0 = 0.8$ for the on-off detectors, and $p = \frac{\pi}{2}$.}
    \label{fig:logneg_alpha_final_state}
\end{figure}

In Fig.~\ref{fig:logneg_alpha_final_state} we plot the logarithmic negativity of the state $\rho_{a_1b_1}$ as a function of the transmission distance for different values of the coherent amplitude $\alpha$. 
We assume that the transmittance of both the channels is given by $T_A$ and $T_B$, respectively, such that, $T_A = 10^{-l \frac{L_{AC}}{10}}$ and $T_B = 10^{-l \frac{L_{BC}}{10}}$, where $l = 0.2$ dB/km is the standard channel loss for telecommunication wavelength~\cite{LCQ12, POS15} and $L_{AC}$ and $L_{BC}$ are the transmission distances between Alice-Charlie and Bob-Charlie respectively. 
To simplify the scenario, we also assume that Charlie is midway between Alice and Bob such that $L_{AC} = L_{BC} = L/2$ such that the total transmission distance is $L$. 
We find that the entanglement of the final state decreases exponentially with the total transmission distance. 
As an example, at $L = 100$ km the logarithmic negativity is $1.3 \times 10^{-4}$ for coherent amplitude $\alpha = 0.6$.

Next, in Fig.~\ref{fig:logneg_length_final_state} we plot the logarithmic negativity as a function of the coherent amplitude $\alpha$ for different transmission distances. 
As it is evident from the Fig. that the shared-entanglement varies non-monotonically on the coherent amplitude ($\alpha$).
We observe that as the transmission distance increases, the optimal value of $\alpha$ becomes less than unity.
For higher transmission distance ($L\geq 150$ Km) this optimal value becomes close to $\alpha=0.5$ (not shown in the Fig.). 
It is found that there also exists an optimal value of $\alpha$ that offers maximum entanglement at a given distance which may be different than the optimal value of $\alpha$ which maximizes the entanglement of the original HE state (as shown in Fig.~\ref{fig:logneg}).

\section{Quantum key distribution using HE states}
\label{sec:qkd}

In this section we benchmark the quality of the shared entangled state in terms of an EB QKD protocol which we set up around the scheme presented in Sec.~\ref{sec:entanglement_sharing}. 
We also consider an eavesdropper, Eve, who may collaborate with Charlie to determine the secure key that is being shared between Alice and Bob.
Additionally, in our protocol we make the following assumptions:
\begin{enumerate}
    \item We assume that Alice and Bob have access to secure laboratories in which they can perform well characterized measurements. 
    Moreover, the measurement devices of Alice and Bob are assumed to be immune to any side-channel attack since no unwanted system may enter or exit the secure laboratories. 
    In the protocol, the DV modes $a_1$ and $b_1$ with Alice and Bob, respectively, are assumed to be in these secure laboratories and do not directly take part in the transmission. 
    On the other hand, the CV modes $a_2$ and $b_2$ are not assumed to be in secure laboratories and as such are vulnerable to eavesdropping attacks. 

    \item We also assume that the quantum channels between Alice-Charlie and Bob-Charlie are characterized by transmission losses only, with no thermal noise. 
    We justify this assumption by demonstrating, in Appendix.~\ref{subsec:hes_fid_charlie_channel}, the scenario with no thermal noise approximates the scenario with some practical value of the same with more than $98\%$ fidelity. 
    This assumption is only required to manage the calculation complexity of evaluating the final DV state between Alice and Bob.

    \item We also consider a third party, Charlie, who is assumed to be untrusted and can collaborate with an eavesdropper, Eve. 
    In the worst case scenario, we assume that he is identified as Eve herself. 
    The QKD protocol, as described in the main text, dictates that Charlie performs certain measurements and publicly declare the outcomes so that Alice and Bob can share an entangled state. 
    In principle, as an eavesdropper, we assume that Charlie may not perform the operations as dictated by the protocol. 
    However, it is required for him to supply some outcomes to the specified measurements to activate the correlations between Alice and Bob. 
    However, if these outcomes are tampered with or even fabricated, the correlations between Alice and Bob will decrease. 
    It is then possible for Alice and Bob to detect the presence of Eve by various methods including state tomography since the parties know the final state they should potentially share. More details on this assumption and the concept of secure laboratories can be found in Ref.~\cite{BP12}.
\end{enumerate}

\subsection{Steps in evaluating key rate}
It should be noted that the steps of the QKD protocol directly follow after step~$4$ in Sec.~\ref{sec:entanglement_sharing} as

\textbf{Step 5:} For the case in which Alice and Bob share $\rho_{a_1b_1}$, they perform two-outcome measurements $\mathcal{M}_A$ and $\mathcal{M}_B$ on their respective subsystems to generate a raw key. 
The choice of measurements is made prior to starting the protocol and the information about this choice is usually publicly available. 
In our protocol, they perform Pauli measurements corresponding to $\sigma_Z$ on their respective subsystems to generate a raw key. 
The length of the raw key that the parties can generate is quantified by the mutual information $I(A:B)$ between them for the observable $\sigma_Z$.

\textbf{Step 6:} Alice and Bob then estimate the amount of information that an adversary, Eve, can have on their raw key. 
This information is quantified by the Holevo bound $\chi(A:E)$ between Alice and Eve. 
In our protocol we consider the Holevo bound to quantify the knowledge about the coherent amplitude $\alpha$, results of the on-off and homodyne measurement which are publicly declared and are actively used in generating the final state between Alice and Bob. 
These results can be used by Eve and as such must be taken care of in the security analysis. 

\subsection{Simulation results on the secured key rate}
Our protocol comprises of two quantum channels: one between Alice and Charlie and another between Bob and Charlie. 
As before, we consider that the transmittance of both the channels is given by $T_A$ and $T_B$, respectively, such that, $T_A = 10^{-l \frac{L_{AC}}{10}}$ and $T_B = 10^{-l \frac{L_{BC}}{10}}$, where $l = 0.2$ dB/km is the standard channel loss for telecom wavelength~\cite{LCQ12, POS15} and $L_{AC}=L_{BC}=L/2$ are the transmission distances between Alice-Charlie and Bob-Charlie respectively such that the total transmission distance is $L$. 

Moreover, we consider that the detectors of Alice and Bob have efficiency $\eta_d$ such that the error rate is given as $Q = 1 - \eta_d$.
With these inefficient detectors, the final secure key rate (See Appendix~\ref{append_sec:calculations_keyrate} for a detailed analysis) for the state given in Eq.~\eqref{eq:final_state} is given as
\begin{align}
    &r \geq P_0 \args{I(A:B) - \chi(A:E)}
    \nonumber 
    \\
    &= P_0 \left\lbrace 1 -\eta_d + \frac{1}{2}\big[ (1+h)\log_2(1+h) + (1-h)\log_2(1-h) \big] 
    \right.
    \nonumber 
    \\
    &~~\left. - \frac{1}{2} \big[ (2-\eta_d)\log_2(2-\eta_d) - (1-\eta_d)\log_2(1-\eta_d) \big]
    \right\rbrace
    \label{eq:secure_key_error},
\end{align}
where it can be seen that the secure key rate only depends on the parameters $h$ (from Eq.~\eqref{eq:final_state}), the detector efficiency of Alice and Bob and the probability with which the final state is prepared. 

\begin{figure}
    \centering
    \includegraphics[angle = -90, scale = 0.35]{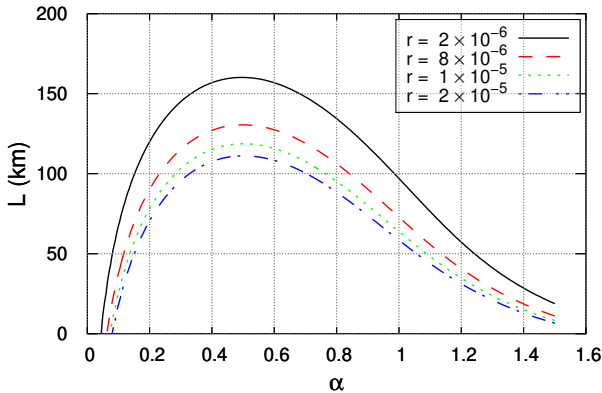}
    \caption{The total transmission distance as a function of the coherent amplitude for different values of the secure key rate. We fix the channel loss at $l = 0.2$ dB/km which corresponds to losses in standard optical fibres. We also fix $\eta_d = 0$ for this analysis. The optimal value of $\alpha$ which maximizes the transmission distance is found to be the same in each case. The unit for the secure key rate $r$ is bits/pulse.}
    \label{fig:keyrate_l_alpha_contour}
\end{figure}

Generally, for an experimental realization of the QKD protocol, the labs of Alice and Bob are fixed at some distance $L$. 
As seen in the main text, $\alpha$ cannot be chosen arbitrarily, as there exists an optimal value which can either maximize the key rate or the total transmission distance. 
In Fig.~\ref{fig:keyrate_l_alpha_contour}, we plot the maximum transmission distance as a function of the coherent amplitude for various values of secure key rate with ideal detector $\eta_d=0$. 
We observe that there exists an optimal value $\alpha \approx 0.5$ which maximizes the total transmission distance for any value of the secure key rate. 

\begin{figure}
    \centering
    \includegraphics[scale = 0.35, angle = -90]{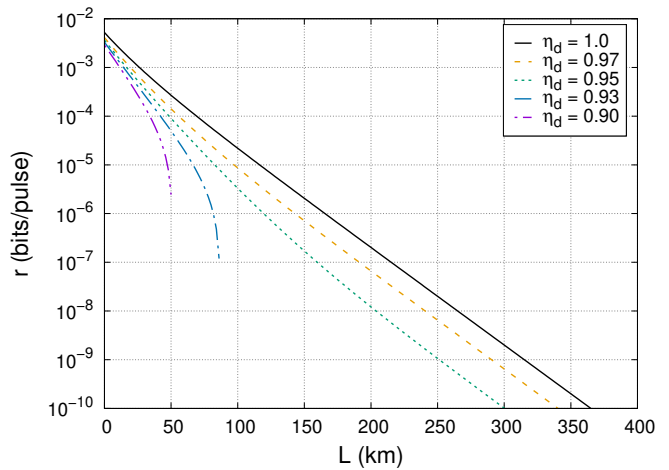}
    \caption{Secure key rate as a function of total transmission distance $L$ for different values of $\eta_d$ in the optimal case, i.e., $\alpha=0.5$. 
    We assume detection efficiencies $\eta_h = 0.55$ for the homodyne detectors, $\eta_0 = 0.8$ for the on-off detectors, and $p = \frac{\pi}{2}$.}
    \label{fig:keyrate_vs_length_error}
\end{figure}

In Fig. \ref{fig:keyrate_vs_length_error} we plot the secure key rate as a function of the total transmission distance $L$ for different values of $\eta_d$. 
We choose the parameters $\eta_h = 0.55$, $\eta_0 = 0.8$, and $p = \frac{\pi}{2}$ to be as realistic as possible and simulate the results for a standard telecom fiber with $l = 0.2$dB/km. 
Furthermore, the value of the coherent amplitude $\alpha$ is chosen to maximize the secure key rate over long transmission distances instead of entanglement. 
For our analysis we choose $\alpha = 0.5$ to optimize the total distance. 
It is also approximately the same value that optimizes the logarithmic negativity of an HE state when its CV part undergoes high photon loss. 
It is seen that under lower errors on Alice's and Bob's side ($\eta_d = 0.97$ and $0.95$), a secure key rate can be achieved for transmission distances around $300$ km indicating that the resultant entangled state is useful. 
However, the maximum achievable distance drastically falls off as $\eta_d$ is increased up to $\eta_d = 0.90$.

\section{Discussion and Conclusion}
\label{sec:conc}
We have shown that HE states between CV and DV systems provide a robust practical solution to the problem of achieving long-distance high-rate entanglement. 
Both requirements are fundamental for a number of applications. 
In this paper we have bench marked the usefulness of the prepared entangled state using an EB-QKD protocol, as it is both a fundamental application and a multipurpose test bed.
In an EB-QKD setup, our results indicate that HE states bring forth the best of both CV and DV systems, resulting in a secure key rate of $\sim 10^{-9}$ bits/pulse at a distance of $250$ km with $5\%$ detection in-efficiency.
This, in itself, represents a significant contribution. 
All this, without using an ultra low-loss fibre (with channel loss $l = 0.16$ dB/km at
$1550$ nm \cite{YCY16}), which allowed an earlier result to achieve transmission distances higher than $400$ km. 
With such a fibre, our approach would allow us to achieve $\sim 10^{-10}$ bits/pulse at a distance of $300$ km for $\eta_d = 0.95$. 

In our analysis, it should be noted that, for the sake of simplicity, we do not consider any thermal noise in the channels. 
However, one can qualitatively show that incorporating a practical value for such thermal noise will not significantly affect our results (see Appendix \ref{append_sec:hes_fid_loss_noise} for detail). 
We leave the detailed quantitative analysis in the presence of thermal noise for future works and acknowledge that it involves lengthy analytical calculations that may have insignificant impact on our findings.

The feasibility of our protocol relies in the fact that HE states with small coherent amplitudes ($\alpha < 1$) can be generated in the lab by several non-deterministic techniques~\cite{JZK14, MHL14} and the generation rate is comparable to the rate of entangled photon pairs in parametric down conversion setups.
As an example, it is possible to prepare HE states where CV and DV parts correspond to photon number state and coherent state, respectively, with fidelity $\approx 0.75$ for $\alpha = 0.5$~\cite{JZK14}. 
While the rate of generation in the source is comparable to that of parametric down conversion sources, losses during transmission are reduced, so the effective rate at destination increases. The fidelity of the preparation could be a limiting factor. 
However, this can be mitigated by using other forms of HE states, most notably with the CV and DV modes corresponding to cat states and polarization, respectively, which offer exceptionally good fidelity of preparation as well as rate of generation~\cite{KJ15, LYH18, PA19, HJM19, WNQ21}. 
We also note that a recent result indicates that it is also possible to deterministically generate HE states with high fidelity~\cite{SM23}

However, it should be noted that there will be effects from phase modulations and phase mismatch in a practical implementation of our scheme.
Commercially available lasers, used in generation of HE states, generally do not have well defined phase stabilization while the optical fibres, used for transmission, may introduce non-linear effects on the signals.
This causes additional concerns for phase-locking and phase-tracking to ensure successful interference at Charlie's end.
Although such issues have been managed in the context of twin-field (TF) QKD \cite{LWL23}, it remains unclear to us whether a similar architecture can be useful in our setup as well and we leave it as an open avenue for future discussions.

Our results highlight the significance of HE states as a resource in high-rate remote entanglement generation, which plays a crucial role in enhancing many quantum information processing tasks such as quantum internet~\cite{TYZ16, WEH18}, quantum digital signature~\cite{YFL17, ZSS21}, and network steering~\cite{JSU21}. 
We believe that our scheme has the potential to drive a new generation of experimental developments in quantum information technology.

\begin{acknowledgements}
The authors thank Stefano Pirandola and Hua-Lei Yin for their comments on an earlier version of the manuscript. This work is supported by QuantERA
grant SECRET, by \href{10.13039/501100011033}{MCINN/AEI} (Project
No.\ PCI2019-111885-2), National Research Foundation of Korea (NRF) grants funded by the Korean government (Grant~Nos.~NRF-2022M3K4A1097117 and NRF-2023R1A2C1006115) via the Institute of Applied Physics at Seoul National University, and by the Institute of Information \& Communications Technology Planning $\&$ Evaluation (IITP) grant funded by the Korea government (MSIT) (IITP-2021-0-01059 and IITP-2023-2020-0-01606). 
AC is also supported by \href{10.13039/501100011033}{MCINN/AEI} (Project No. PID2020-113738GB-I00). J.~S. also acknowledges the support from the National Science and Technological Council (NSTC) of Taiwan through grant no. NSTC 112-2628-M-006-007-MY4 and NSTC 112-2811-M-006-033-MY4.
\end{acknowledgements}


\newpage
\appendix 
\section{Generation of the hybrid entangled state}
\label{append_sec:hesgeneration}

In this section we outline a process, using a setup in line with Ref.~\cite{JZK14}, that can be used to experimentally generate a hybrid entangled (HE) state of the form 
\begin{equation}
  \ket{\psi}_{ab}=\frac{1}{\sqrt{2}}\left( \ket{0}_{a}\ket{\alpha}_{b} + \ket{1}_{a}\ket{-\alpha}_{b} \right)
  \label{suppl_eq:hes_def}
\end{equation}
between modes $a$ and $b$, where $\ket{0}$ and $\ket{1}$ correspond to photon number states, and $\ket{\alpha}$ is the coherent state with coherent amplitude $\alpha$. 

$\ket{n}$ and $\ket{\alpha}$ correspond to, respectively, the energy eigenstate and coherent state of a quantized electromagnetic field, where $n$ is the number of photons in the state. It is possible to realize the energy eigenstates as a single-photon qubit by only considering the photon number states corresponding to $\ket{0}$ and $\ket{1}$. This is our discrete variable (DV) system and the multiphoton coherent state is our continuous variable (CV) system.
The key idea of HE state generation hinges on conditional photon addition and erasing the path information of photon addition.
There are several ways of achieving photon addition. This includes a model which uses single photon sources and beam splitters (BS) and another model which uses a parametric-down-converter (PDC) with a weak pump. Since the BS setup and the PDC are equivalent \cite{BPJ99}, here we use the BS model for photon addition.
The following is a step-by-step description of the generation of the HE state in Fig.~$1$(b) of the main text.

{\bf Step 1:} A vacuum state $\ket{0}$ in mode $a$ is mixed with a single-photon state $\ket{1}$ in mode $c$ using a BS (BS$_1$) with transmittance $T$. 
Similarly, a coherent state $\ket{\alpha}$ in mode $c$ is mixed with another single-photon state in mode $d$ using another BS (BS$_2$) with transmittance $T$. 
The output states from each of these two BSs are
\begin{align}
    \ket{\psi}_{ac}^{\rm{BS}_1} &= \sqrt{1-T}\ket{1}_a\ket{0}_c + \sqrt{T}\ket{0}_a\ket{1}_c
    \nonumber 
    \\
    \ket{\psi}_{bd}^{\rm{BS}_2} &= \sqrt{1-T} \hat{b}^{\dagger}\ket{\alpha}_b\ket{0}_d + \sqrt{T}\ket{\alpha}_b\ket{1}_d,
\end{align}
where $\hat{b}^\dagger$ is the creation operator acting on mode $b$ and $\ket{\psi}_{ac}^{\rm{BS}_1}$ is the output state from BS$_1$, while $\ket{\psi}_{bd}^{\rm{BS}_2}$ is the output state from BS$_2$. 
The BS transmittance $T$ can be fine-tuned according to experimental requirements to yield maximum probability for photon addition.
Therefore, the $4$-mode state at the output of BS$_1$ and BS$_2$ is
\begin{widetext}
    \begin{align}
        \ket{\psi}_{ab,cd}^{\rm{BS}_{1,2}} &= \sqrt{T(1-T)} \left( \ket{1}_a \ket{\alpha}_b \otimes \ket{0}_c\ket{1}_d
        +
        \ket{0}_a \hat{b}^{\dagger}\ket{\alpha}_b \otimes \ket{1}_c\ket{0}_d
        \right)
        +(1-T)\ket{1}_a \hat{b}^{\dagger}\ket{\alpha}_b \otimes \ket{0}_c\ket{0}_d
        +
        T\ket{0}_a\ket{\alpha}_b \otimes \ket{1}_c\ket{1}_d.
    \end{align}
\end{widetext}

{\bf Step 2:} The outgoing single-photon modes (shown by gray dashed-lines in Fig.~$1$(b) of the main text) from both BS$_1$ and BS$_2$ are mixed with each other using a another BS (BS$_3$) with transmittance $\tau$.
This leads to a $4$-mode state at the output of BS$_3$ which can be written as
\begin{widetext}
    \begin{align}
        \ket{\psi}_{ab,cd}^{\rm{BS}_3} &= \sqrt{T(1-T)} \left[ \ket{1}_a \ket{\alpha}_b \otimes \Big( -\sqrt{1-\tau}\ket{0}_c\ket{1}_d + \sqrt{\tau}\ket{1}_c\ket{0}_d \Big)
        +
        \ket{0}_a \hat{b}^{\dagger} \ket{\alpha}_b \otimes \Big( \sqrt{1-\tau}\ket{1}_c\ket{0}_d + \sqrt{\tau} \ket{0}_c\ket{1}_d \Big)
        \right]
        \nonumber 
        \\
        &~~~ +(1-T)\ket{1}_a \hat{b}^{\dagger}\ket{\alpha}_b \otimes \ket{0}_c\ket{0}_d
        +
        T \ket{0}_a \ket{\alpha}_b \otimes \Big( \sqrt{1-\tau}\ket{2}_c\ket{0}_d + \sqrt{\tau}\ket{0}_c\ket{2}_d \Big).
    \end{align}
\end{widetext}  

{\bf Step 3:} We now detect the output modes of BS${}_3$ via single-photon detectors D$_1$ and D$_2$.
Since the total photon number at the output of BS$_3$ is $1$, it indicates that both D$_1$ and D$_2$ cannot click simultaneously. 
We post-select the state when only the detector D$_1$ clicks and discard the runs whenever the detector D$_2$ clicks.
After post-selection, the state between modes $a$ and $b$ is
\begin{align}
    \ket{\psi}_{ab}^{\rm{D}_1} &= \bra{1}_c\bra{0}_d \ket{\psi}_{ab,cd}^{\rm{BS}_3}
    \nonumber 
    \\
    &= \sqrt{T(1-T)} \left( \sqrt{\tau}\ket{1}_a\ket{\alpha}_b + \sqrt{1-\tau} \ket{0}_a \hat{b}^{\dagger}\ket{\alpha}_b \right).
\end{align}
  
We can now use the fact that $n$-photon-added coherent state is a good approximation to another coherent state with amplified amplitude, i.e., $\frac{\hat{b}^{\dagger n}}{\sqrt{N}} \ket{\alpha} \approx \ket{g\alpha}$ \cite{JZK14}, where $N$ is the corresponding normalization constant and $g \geq 1$ is the amplification factor. 
This leads to the result $\hat{b}^{\dagger}\ket{\alpha}_b \approx \frac{1}{\sqrt{1-\alpha^2}}\ket{g\alpha}_b$, where $g$ is properly chosen.
Then, by setting $\tau = \frac{1+\alpha^2}{2+\alpha^2}$ and using the approximation we get,
\begin{equation}
 \ket{\psi}_{ab}^{\rm{D}_1} \approx \sqrt{\frac{T(1-T)}{2+\alpha^2}} \Bigl( \ket{1}_a\ket{\alpha}_b + \ket{0}_a\ket{g\alpha}_b \Bigr).
\end{equation}
  
{\bf Step 4:} Next, we displace the mode $b$ by performing a displacement operator on this mode given by $D_b\left(- \frac{\alpha+g\alpha}{2}\right) = \exp \left[ - \frac{\alpha+g\alpha}{2} (\hat{b}^{\dagger} - \hat{b})\right]$, where $\hat{b}$ is the annihilation operator.
This leads to the final normalized HE state
\begin{equation}
\ket{\psi}_{ab} = \frac{1}{\sqrt{2}} \left( \ket{0}_a\ket{\alpha_f}_b + \ket{1}_a\ket{-\alpha_f}_b \right),
\label{eq:hes}
\end{equation}
where $\alpha_f= \frac{(g-1)\alpha}{2}$.

\section{Shared entangled state between Alice and Bob}
\label{append_sec:final state}

In this section, we calculate the the state obtained after performing the entanglement swapping operation by Charlie.
We also calculate the states obtained after every step of the protocol starting from the initial resource of HE states. 
The steps of the protocol are detailed in the main manuscript.

\subsection{Initial states and channel transmission} 
\label{subsec:preli}


We denote the two hybrid entangled states with Alice and Bob as
\begin{align}
    \ket{\psi}_{a_1a_2} &= \frac{1}{\sqrt{2}}\left( \ket{0}_{a_1}\ket{\alpha}_{a_2} + \ket{1}_{a_1}\ket{-\alpha}_{a_2} \right)
    \nonumber 
    \\
    \ket{\psi}_{b_1b_2} &= \frac{1}{\sqrt{2}}\left( \ket{0}_{b_1}\ket{\alpha}_{b_2} + \ket{1}_{b_1}\ket{-\alpha}_{b_2} \right),
\end{align}
respectively. The initial $4$-mode resource state can be written as
\begin{align}
    &\ket{\psi}_{\substack{a_1a_2\\b_1b_2}} = \ket{\psi}_{a_1a_2} \ket{\psi}_{b_1b_2}
    \nonumber 
    \\
    &= \frac{1}{2} \left( 
    \ket{00}_{a_1b_1} \ket{\alpha}_{a_2}\ket{\alpha}_{b_2} + 
    \ket{11}_{a_1b_1} \ket{-\alpha}_{a_2}\ket{-\alpha}_{b_2}
    \right.
    \nonumber 
    \\
    &~~\left.+ 
    \ket{01}_{a_1b_1} \ket{\alpha}_{a_2}\ket{-\alpha}_{b_2} + 
    \ket{10}_{a_1b_1} \ket{-\alpha}_{a_2}\ket{\alpha}_{b_2}
    \right),
\end{align}
where $\ket{ij}_{a_1b_1}=\ket{i}_{a_1}\ket{j}_{b_1}$ $\forall i,j\in \lbrace 0,1\rbrace$.

Alice and Bob both send their multiphoton part (modes $a_2$ and $b_2$) to a third distant party Charlie for mixing and subsequent measurements through a noisy/lossy channel with transmittance $T$.  
Such channels could be modelled in terms of an effective beam splitter (BS) with transmittance $T$, where the input state is fed at one of the inputs of the BS while the other input is initialised as a vacuum state.
The action of a BS with transmittance $T$ on the input modes is given by a unitary $U^{ab}_T$ implementing the following transformation:
\begin{equation}
\begin{pmatrix} 
\hat{a} \\
\hat{b}
\end{pmatrix}\rightarrow
\begin{pmatrix} 
\hat{a}' \\
\hat{b}'
\end{pmatrix} = 
\begin{pmatrix} 
\sqrt{T} & \sqrt{1-T} \\
-\sqrt{1-T} & \sqrt{T}
\end{pmatrix}
\begin{pmatrix} 
\hat{a} \\
\hat{b}
\end{pmatrix}.
\end{equation}
$T = \frac{1}{2}$ corresponds to a balanced ($50:50$) BS.
As a consequence, the action of the channel on a coherent state ($\ket\alpha$) in mode $a$ is described as $U^{ab}_T\ket\alpha_{a}\otimes \ket 0_b\rightarrow \ket\alpha_{a^{'}}\otimes \ket 0_{b^{'}} = \ket{\sqrt{T}\alpha}_a\otimes \ket{\sqrt{1-T}\alpha}_b$, where $U^{ab}_T$ is the corresponding BS unitary operation.
Subsequently, the resultant state is obtained by tracing over the ancillary mode $b$.

Similarly, the noisy transmission of modes $a_2$ and $b_2$ could be described by using two BSs with transmittance $T$, each one in the paths of modes $a_2$ and $b_2$ with ancillary modes given by $f_a$ and $f_b$, respectively.
The resultant noisy/lossy state is obtained by tracing over the ancillary modes ($f_a$ and $f_b$).
Therefore, the total input state to Charlie before mixing is
\begin{widetext}
    \begin{align}
        &\ket{\psi}_{\substack{a_1,b_1;a_2^{'},b_2^{'}\\ f_a^{'},f_b^{'}}} = U_{T}^{(a_2,f_a)} \otimes U_{T}^{(b_2,f_b)} \ket{\psi}_{\substack{a_1a_2\\b_1b_2}} \otimes \ket{0}_{f_a} \ket{0}_{f_b}
        \nonumber
        \\
        &=\frac{1}{2} \left( 
        \ket{00}_{a_1b_1} \ket{\sqrt{1-T}\alpha}_{f_a} \ket{\sqrt{1-T}\alpha}_{f_b} \ket{\sqrt{T}\alpha}_{a_2}\ket{\sqrt{T}\alpha}_{b_2}
        +
        \ket{11}_{a_1b_1} \ket{-\sqrt{1-T}\alpha}_{f_a} \ket{-\sqrt{1-T}\alpha}_{f_b} \ket{-\sqrt{T}\alpha}_{a_2}\ket{-\sqrt{T}\alpha}_{b_2}
        \right.
        \nonumber
        \\
        &~~ +\left.
        \ket{01}_{a_1b_1} \ket{\sqrt{1-T}\alpha}_{f_a} \ket{-\sqrt{1-T}\alpha}_{f_b} \ket{\sqrt{T}\alpha}_{a_2}\ket{-\sqrt{T}\alpha}_{b_2}
        +
        \ket{10}_{a_1b_1} \ket{-\sqrt{1-T}\alpha}_{f_a} \ket{\sqrt{1-T}\alpha}_{f_b} \ket{-\sqrt{T}\alpha}_{a_2}\ket{\sqrt{T}\alpha}_{b_2}
        \right),
    \end{align}
\end{widetext}
where $U_{T}^{(a_2,f_a)}$ and $U_{T}^{(b_2,f_b)}$ are the BS unitary operations corresponding to the respective channels with transmittance $T$.
Charlie now mixes the incoming multiphoton modes ($a_2$ and $b_2$) through a balanced BS ($\rm{BS}_1$) leading to the four mode entangled state
\begin{widetext}
    \begin{align}
        &\ket{\psi}^{\rm{BS}_1}_{\substack{a_1,b_1;a_2^{''},b_2^{''}\\ f_a^{'},f_b^{'}}} = U^{(a_2,b_2)}_{\rm{BS}_1} \ket{\psi}_{\substack{a_1,b_1;a_2^{'},b_2^{'}\\ f_a^{'},f_b^{'}}}
        \nonumber
        \\
        &= \frac{1}{2} \left( 
        \ket{00}_{a_1b_1} \ket{\sqrt{1-T}\alpha}_{f_a} \ket{\sqrt{1-T}\alpha}_{f_b} \ket{0}_{b_2}\ket{\sqrt{2T}\alpha}_{a_2}
        +
        \ket{11}_{a_1b_1} \ket{-\sqrt{1-T}\alpha}_{f_a} \ket{-\sqrt{1-T}\alpha}_{f_b} \ket{0}_{b_2^{'}}\ket{-\sqrt{2T}\alpha}_{a_2}
        \right.
        \nonumber
        \\
        &~~\left. +
        \ket{01}_{a_1b_1} \ket{\sqrt{1-T}\alpha}_{f_a} \ket{-\sqrt{1-T}\alpha}_{f_b} \ket{\sqrt{2T}\alpha}_{b_2} \ket{0}_{a_2}
        +
        \ket{10}_{a_1b_1} \ket{-\sqrt{1-T}\alpha}_{f_a} \ket{\sqrt{1-T}\alpha}_{f_b} \ket{-\sqrt{2T}\alpha}_{b_2} \ket{0}_{a_2}
        \right).
        \label{append_eq:psibs1}
    \end{align}
\end{widetext}

It can be seen from Eq.~(\ref{append_eq:psibs1}) that, in the total $4$-mode entangled state after mixing by Charlie, the vacuum state contribution in mode $a_2$ appears with probability $1/2$.
Our primary aim is to postselect the state \eqref{append_eq:psibs1} in $\ket{0}_{a_2}$.


\subsection{State after the on-off measurement}
\label{subsec:rhoonoff}


The additional coherent state sent by Bob to Charlie $\left( \ket{\sqrt{2}\alpha} \right)$ becomes $\ket{\sqrt{2T}\alpha}$ as a result of transmission through lossy channel.
As is described in \cite{JZK14}, for this purpose Charlie first mixes the outgoing $a_2$ mode with this additional state $\left( \ket{\sqrt{2T}\alpha} \right)$ in mode $c$ through the second balanced beam splitter ($\rm{BS}_2$).
Consequently, the state after the mixing at $\rm{BS}_2$ is given by
\begin{widetext}
    \begin{align}
        \ket{\psi}^{\rm{BS}_2}_{\substack{a_1,b_1;b_2^{''}\\ f_a^{'},f_b^{'} \\ a_2^{'''},c^{'}}} &= U^{(a_2,c)}_{\rm{BS}_1} \ket{\psi}^{\rm{BS}_1}_{\substack{a_1,b_1;a_2^{''},b_2^{''}\\ f_a^{'},f_b^{'}}} \otimes \ket{\sqrt{2T}\alpha}_c
        \nonumber
        \\
        &= \frac{1}{2} \left( 
        \ket{00}_{a_1b_1} \ket{\sqrt{1-T}\alpha}_{f_a} \ket{\sqrt{1-T}\alpha}_{f_b} \ket{0}_{b_2} \ket{2\sqrt{T}\alpha}_{a_2}\ket{0}_c
        \right.
        \nonumber
        \\
        &~~+\left.
        \ket{11}_{a_1b_1} \ket{-\sqrt{1-T}\alpha}_{f_a} \ket{-\sqrt{1-T}\alpha}_{f_b} \ket{0}_{b_2} \ket{0}_{a_2}\ket{-2\sqrt{T}\alpha}_c
        \right.
        \nonumber
        \\
        &~~+\left.
        \ket{01}_{a_1b_1} \ket{\sqrt{1-T}\alpha}_{f_a} \ket{-\sqrt{1-T}\alpha}_{f_b} \ket{\sqrt{2T}\alpha}_{b_2} \ket{\sqrt{T}\alpha}_{a_2}\ket{-\sqrt{T}\alpha}_c
        \right.
        \nonumber
        \\
        &~~+\left.
        \ket{10}_{a_1b_1} \ket{-\sqrt{1-T}\alpha}_{f_a} \ket{\sqrt{1-T}\alpha}_{f_b} \ket{-\sqrt{2T}\alpha}_{b_2} \ket{\sqrt{T}\alpha}_{a_2}\ket{-\sqrt{T}\alpha}_c
        \right).
        \label{append_eq:psibs2}
    \end{align}
\end{widetext}

As it can be seen from Eq.~(\ref{append_eq:psibs2}), if both the detectors at the output of $\rm{BS}_2$ click then the contribution can arise only from the respective part in  \eqref{append_eq:psibs2}, i.e., from the part containing $\ket{0}_{a_2}$.
Experimentally, this could be achieved unambiguously by performing the operation $\Pi_0=(\mathbf{I}-\ket 0\bra 0) \otimes (\mathbf{I}-\ket 0\bra 0)$ on \eqref{append_eq:psibs2} using two {\em on-off} detectors at both the output ports of $\rm{BS}_2$. 

However, here we consider {\em non-ideal} detectors with efficiency $\eta_{o}$ ($0 \leq \eta_{o} \leq 1$).
Similar to the case of transmission channels, this could be analysed by considering two additional BS with transmittance $\eta_{o}$ and two ancillary modes $g_a$ and $g_c$ for modes $a_2$ and $c$, respectively.
Therefore, before the {\em on-off} detectors, the total state is given by
\begin{widetext}
    \begin{align}
        \ket{\psi}^{tot}_{\substack{a_1,b_1;b_2^{''}\\ f_a^{'},f_b^{'},g_a^{'},g_c^{'} \\ a_2^{''''},c^{''}}} &= U^{(a_2,g_a)}_{\eta_{o}} \otimes U^{(c,g_c)}_{\eta_{o}} \ket{\psi}^{\rm{BS}_2}_{\substack{a_1,b_1;b_2^{''}\\ f_a^{'},f_b^{'} \\ a_2^{'''},c^{'}}} \otimes \ket{0}_{g_a} \ket{0}_{g_c}
        \nonumber
        \\
        &= \frac{1}{2} \left[
        \ket{00}_{a_1b_1} 
        \ket{0}_{b_2^{''}} 
        \ket{\sqrt{1-T}\alpha}_{f_a} \ket{\sqrt{1-T}\alpha}_{f_b} 
        \ket{2\sqrt{T(1-\eta_{o})}\alpha}_{g_a} \ket{0}_{g_c}
        \ket{2\sqrt{T\eta_{o}}\alpha}_{a_2}\ket{0}_c
        \right.
        \nonumber
        \\
        &~~+\left.
        \ket{11}_{a_1b_1} 
        \ket{0}_{b_2}
        \ket{-\sqrt{1-T}\alpha}_{f_a} \ket{-\sqrt{1-T}\alpha}_{f_b} 
        \ket{0}_{g_a} \ket{-2\sqrt{T(1-\eta_o)}\alpha}_{g_c} 
        \ket{0}_{a_2}\ket{-2\sqrt{T\eta_o}\alpha}_c
        \right.
        \nonumber
        \\
        &~~+\left(
        \ket{01}_{a_1b_1} 
        \ket{\sqrt{2T}\alpha}_{b_2}
        \ket{\sqrt{1-T}\alpha}_{f_a} \ket{-\sqrt{1-T}\alpha}_{f_b} 
        +
        \ket{10}_{a_1b_1} 
        \ket{-\sqrt{2T}\alpha}_{b_2} 
        \ket{-\sqrt{1-T}\alpha}_{f_a} \ket{\sqrt{1-T}\alpha}_{f_b} 
        \right)
        \nonumber 
        \\
        &~~~~\left.
        \ket{\sqrt{T(1-\eta_o)}\alpha}_{g_a} \ket{-\sqrt{T(1-\eta_o)}\alpha}_{g_c}
        \ket{\sqrt{T\eta_0}\alpha}_{a_2}\ket{-\sqrt{T\eta_o}\alpha}_{c}
        \right]
        \nonumber 
        \\
        &= \frac{1}{2} \left[ 
        \ket{00}_{a_1b_1} 
        \ket{0}_{b_2} 
        \ket{\sqrt{T'}\alpha}_{f_a} \ket{\sqrt{T'}\alpha}_{f_b} 
        \ket{2\sqrt{T\eta_o'}\alpha}_{g_a} \ket{0}_{g_c} 
        \ket{2\sqrt{T\eta_o}\alpha}_{a_2}\ket{0}_c
        \right.
        \nonumber
        \\
        &~~+\left.
        \ket{11}_{a_1b_1} 
        \ket{0}_{b_2}
        \ket{-\sqrt{T'}\alpha}_{f_a} \ket{-\sqrt{T'}\alpha}_{f_b} 
        \ket{0}_{g_a} \ket{-2\sqrt{T\eta_o'}\alpha}_{g_c} 
        \ket{0}_{a_2}\ket{-2\sqrt{T\eta_o}\alpha}_c
        \right.
        \nonumber
        \\
        &~~+\left(
        \ket{01}_{a_1b_1} 
        \ket{\sqrt{2T}\alpha}_{b_2}
        \ket{\sqrt{T'}\alpha}_{f_a} \ket{-\sqrt{T'}\alpha}_{f_b} 
        ~+~
        \ket{10}_{a_1b_1} 
        \ket{-\sqrt{2T}\alpha}_{b_2} 
        \ket{-\sqrt{T'}\alpha}_{f_a} \ket{\sqrt{T'}\alpha}_{f_b}
        \right)
        \nonumber
        \\
        &~~\left. 
        \ket{\sqrt{T\eta_o'}\alpha}_{g_a} \ket{-\sqrt{T\eta_o'}\alpha}_{g_c}
        \ket{\sqrt{T\eta_0}\alpha}_{a_2}\ket{-\sqrt{T\eta_o}\alpha}_c
        \right],
        \label{append_eq:psionoff}
    \end{align}
\end{widetext}
where $T'=1-T$ and $\eta_o'=1-\eta_o$.

Charlie is now supposed to make the measurement of $\Pi_0^{a_2,c}=(\mathds{1}-\ket 0\bra 0)_{a_2} \otimes (\mathds{1}-\ket 0\bra 0)_c$ on the state in Eq. \eqref{append_eq:psionoff}.
After the measurement of these operators ($\Pi_0$) the total state collapses to $\rho_{a_1,b_1;b_2}^0={\rm{tr}}_{\substack{f_a,f_b\\g_a,g_c\\a_2,c}} \Big( \ket{\psi_0}\bra{\psi_0} \Big)/N_0$, where 
$\ket{\psi_0}=\Pi_0^{a_2,c} \ket{\psi}^{tot}_{\substack{a_1,b_1;b_2^{''}\\ f_a^{'},f_b^{'},g_a^{'},g_c^{'} \\ a_2^{''''},c^{''}}}$ and the normalization constants are $N_0={\rm{tr}}_{\substack{a_1,b_1,a_2,b_2,c\\f_a,f_b,g_a,g_c}} \Big( \ket{\psi_0}\bra{\psi_0} \Big)$.
It must be noted that the state $\rho_{a_1,b_1;b_2}^0$ is obtained with probability $P_0={\rm{tr}}_{\substack{a_1,b_1\\b_2}} \left( \rho_{a_1,b_1;b_2}^0 \right)$.

Let us look at the result first
\begin{widetext}
    \begin{align}
        \Pi_0^{a_2,c}\ket{\alpha}_{a_2}\ket{\beta}_c &=
        \Big( \mathds{1}_{a_2}\otimes\mathds{1}_c - \ket{0}_{a_2}\bra{0}\otimes \mathds{1}_c - \mathds{1}_{a_2}\otimes \ket{0}_{c}\bra{0} + \ket{0}_{a_2}\bra{0}\otimes \ket{0}_{c}\bra{0} \Big) \ket{\alpha}_{a_2}\ket{\beta}_c
        \nonumber
        \\
        &= \ket{\alpha}_{a_2}\ket{\beta}_c - e^{-\alpha^2/2} \ket{0}_{a_2}\ket{\beta}_c - e^{-\beta^2/2} \ket{\alpha}_{a_2}\ket{0}_c + e^{-(\alpha^2+\beta^2)/2} \ket{0}_{a_2}\ket{0}_c.
        \label{append_eq:pi0ab1}
    \end{align}
    leading to
    \begin{align}
        \Pi_0^{a_2,c}\ket{\alpha}_{a_2}\ket{0}_c &= \Pi_0^{a_2,c}\ket{0}_{a_2}\ket{\alpha}_c = 0 
        \nonumber
        \\
        \Pi_0^{a_2,c}\ket{\alpha}_{a_2}\ket{-\alpha}_c &= 
        \ket{\alpha}_{a_2}\ket{-\alpha}_c - e^{-\alpha^2/2}\ket{0}_{a_2}\ket{-\alpha}_c - e^{-\alpha^2/2}\ket{\alpha}_{a_2}\ket{0}_c + e^{-\alpha^2} \ket{0}_{a_2}\ket{0}_c.
        \label{append_eq:pi0ab2}
    \end{align}

    Deploying the results of Eq. \eqref{append_eq:pi0ab2} in Eq. \eqref{append_eq:psionoff}, we obtain 
    \begin{align}
        &\ket{\psi_0} = \Pi_0^{a_2,c} \ket{\psi}^{tot}_{\substack{a_1,b_1;b_2^{''}\\ f_a^{'},f_b^{'},g_a^{'},g_c^{'} \\ a_2^{''''},c^{''}}}
        \nonumber 
        \\
        &= \frac{1}{2} \left[ 
        \ket{00}_{a_1b_1} 
        \ket{0}_{b_2} 
        \ket{\sqrt{T'}\alpha}_{f_a} \ket{\sqrt{T'}\alpha}_{f_b} 
        \otimes 
        \ket{2\sqrt{T\eta_o'}\alpha}_{g_a} \ket{0}_{g_c} 
        \times 
        0
        \right.
        \nonumber
        \\
        &~~~~+\left.
        \ket{11}_{a_1b_1} 
        \ket{0}_{b_2}
        \ket{-\sqrt{T'}\alpha}_{f_a} \ket{-\sqrt{T'}\alpha}_{f_b} 
        \otimes 
        \ket{0}_{g_a} \ket{-2\sqrt{T\eta_o'}\alpha}_{g_c} 
        \times 
        0
        \right.
        \nonumber
        \\
        &~~+\left(
        \ket{01}_{a_1b_1} 
        \ket{\sqrt{2T}\alpha}_{b_2}
        \ket{\sqrt{T'}\alpha}_{f_a} \ket{-\sqrt{T'}\alpha}_{f_b} 
        +
        \ket{10}_{a_1b_1} 
        \ket{-\sqrt{2T}\alpha}_{b_2} 
        \ket{-\sqrt{T'}\alpha}_{f_a} \ket{\sqrt{T'}\alpha}_{f_b}
        \right)
        \otimes 
        \ket{\sqrt{T\eta_o'}\alpha}_{g_a} \ket{-\sqrt{T\eta_o'}\alpha}_{g_c}
        \otimes
        \nonumber 
        \\
        &~~\left.\left(
        \ket{\sqrt{\eta_oT}\alpha}_{a_2}\ket{-\sqrt{\eta_oT}\alpha}_c - e^{-\eta_oT\alpha^2/2} \ket{0}_{a_2}\ket{-\sqrt{\eta_oT}\alpha}_c - e^{-\eta_oT\alpha^2/2} \ket{\sqrt{\eta_oT}\alpha}_{a_2}\ket{0}_c + e^{-\eta_oT\alpha^2} \ket{0}_{a_2}\ket{0}_c
        \right)\right]
        \nonumber
        \\
        &= \frac{1}{2} \left(
        \ket{01}_{a_1b_1} 
        \ket{\sqrt{2T}\alpha}_{b_2}
        \ket{\sqrt{T'}\alpha}_{f_a} \ket{-\sqrt{T'}\alpha}_{f_b} 
        +
        \ket{10}_{a_1b_1} 
        \ket{-\sqrt{2T}\alpha}_{b_2} 
        \ket{-\sqrt{T'}\alpha}_{f_a} \ket{\sqrt{T'}\alpha}_{f_b}
        \right)
        \otimes 
        \nonumber
        \\
        &~~ \ket{\sqrt{T\eta_o'}\alpha}_{g_a} \ket{-\sqrt{T\eta_o'}\alpha}_{g_c}
        \otimes 
        \ket{\Psi}_{a_2c},
    \end{align}
    where
    \begin{equation}
        \ket{\Psi}_{a_2c} = \ket{\sqrt{\eta_oT}\alpha}_{a_2}\ket{-\sqrt{\eta_oT}\alpha}_c - e^{-\eta_oT\alpha^2/2} \ket{0}_{a_2}\ket{-\sqrt{\eta_oT}\alpha}_c - e^{-\eta_oT\alpha^2/2} \ket{\sqrt{\eta_oT}\alpha}_{a_2}\ket{0}_c + e^{-\eta_oT\alpha^2} \ket{0}_{a_2}\ket{0}_c.
    \end{equation}

    It can be further shown that 
    \begin{align}
        &{\rm{tr}} \Big( \ket{\Psi}_{a_2c}\bra{\Psi} \Big) =
        {\rm{tr}}\left[ \left(
        \bra{\sqrt{\eta_oT}\alpha}\ket{\sqrt{\eta_oT}\alpha} \bra{-\sqrt{\eta_oT}\alpha}\ket{-\sqrt{\eta_oT}\alpha} 
        + 
        e^{-\eta_oT\alpha^2} \bra{0}\ket{0} \bra{-\sqrt{\eta_oT}\alpha}\ket{-\sqrt{\eta_oT}\alpha}
        \right. 
        \right.
        \nonumber
        \\
        &~~+\left. \left.
        e^{-\eta_oT\alpha^2} \bra{\sqrt{\eta_oT}\alpha}\ket{\sqrt{\eta_oT}\alpha} \bra{0}\ket{0}
        + 
        e^{-2\eta_oT\alpha^2} \bra{0}\ket{0} \bra{0}\ket{0}
        \right)
        + 2 \left(
        -e^{-\eta_oT\alpha^2/2} \bra{0}\ket{\sqrt{\eta_oT}\alpha} \bra{-\sqrt{\eta_oT}\alpha}\ket{-\sqrt{\eta_oT}\alpha}
        \right.
        \right. 
        \nonumber
        \\
        &~~ \left.- 
        e^{-\eta_oT\alpha^2/2} \bra{\sqrt{\eta_oT}\alpha}\ket{\sqrt{\eta_oT}\alpha} \bra{0}\ket{-\sqrt{\eta_oT}\alpha}
        +
        e^{-\eta_oT\alpha^2} \bra{0}\ket{\sqrt{\eta_oT}\alpha} \bra{0}\ket{-\sqrt{\eta_oT}\alpha}
        \right) 
        \nonumber
        \\
        &~~+ 2 \left(
        e^{-\eta_oT\alpha^2} \bra{\sqrt{\eta_oT}\alpha}\ket{0} \bra{0}\ket{-\sqrt{\eta_oT}\alpha} 
        -
        e^{-3\eta_oT\alpha^2/2} \bra{0}\ket{0} \bra{0}\ket{-\sqrt{\eta_oT}\alpha}
        \right)
        -~ 
        2 \left.
        e^{-3\eta_oT\alpha^2/2} \bra{0}\ket{\sqrt{\eta_oT}\alpha} \bra{0}\ket{0}
        \right]
        \nonumber
        \\
        &= \left( 1+e^{-\eta_oT\alpha^2}+e^{-\eta_oT\alpha^2}+e^{-2\eta_oT\alpha^2} \right)
        +
        2 \left( -e^{-\eta_oT\alpha^2}-e^{-\eta_oT\alpha^2}+e^{-2\eta_oT\alpha^2} \right)
        +
        2 \left( e^{-2\eta_oT\alpha^2}-e^{-2\eta_oT\alpha^2}\right)
        - 
        2e^{-2\eta_oT\alpha^2}
        \nonumber
        \\
        &=1-2e^{-\eta_oT\alpha^2}+e^{-2\eta_oT\alpha^2} = \left( 1 - e^{-\eta_oT\alpha^2} \right)^2.
    \end{align}

    Hence
    \begin{align}
        &{\rm{tr}}_{\substack{f_a,f_b\\g_a,g_c\\a_2,c}} \Big( \ket{\psi_0}\bra{\psi_0} \Big)
        \nonumber 
        \\
        &= \frac{\left( 1 - e^{-\eta_oT\alpha^2} \right)^2}{4}
        {\rm{tr}}_{f_a,f_b} \left[ \left( 
        \ket{01}_{a_1b_1} \bra{01}
        \otimes
        \ket{\sqrt{2T}\alpha}_{b_2}\bra{\sqrt{2T}\alpha}
        \otimes
        \ket{\sqrt{T'}\alpha}_{f_a}\bra{\sqrt{T'}\alpha} 
        \otimes 
        \ket{-\sqrt{T'}\alpha}_{f_b}\bra{-\sqrt{T'}\alpha}
        \right.
        \right.
        \nonumber 
        \\
        &~~~~~~~~~~~~~~~~~~~~~~~~~~~~~~~~~~ +\ket{10}_{a_1b_1} \bra{10} 
        \otimes
        \ket{-\sqrt{2T}\alpha}_{b_2}\bra{-\sqrt{2T}\alpha}
        \otimes
        \ket{-\sqrt{T'}\alpha}_{f_a}\bra{-\sqrt{T'}\alpha} 
        \otimes 
        \ket{\sqrt{T'}\alpha}_{f_b}\bra{\sqrt{T'}\alpha}
        \nonumber
        \\
        &~~~~~~~~~~~~~~~~~~~~~~~~~~~~~~~~~~ +\ket{01}_{a_1b_1} \bra{10} 
        \otimes
        \ket{\sqrt{2T}\alpha}_{b_2}\bra{-\sqrt{2T}\alpha}
        \otimes 
        \ket{\sqrt{T'}\alpha}_{f_a}\bra{-\sqrt{T'}\alpha} \otimes \ket{-\sqrt{T'}\alpha}_{f_b}\bra{\sqrt{T'}\alpha}
        \nonumber 
        \\
        &~~~~~~~~~~~~~~~~~~~~~~~~~~~~~~~~~~ +\left.\left. \ket{10}_{a_1b_1} \bra{01} 
        \otimes
        \ket{-\sqrt{2T}\alpha}_{b_2}\bra{\sqrt{2T}\alpha}
        \otimes 
        \ket{-\sqrt{T'}\alpha}_{f_a}\bra{\sqrt{T'}\alpha} \otimes \ket{\sqrt{T'}\alpha}_{f_b}\bra{-\sqrt{T'}\alpha}
        \right)
        \right]
        \nonumber
        \\
        &~~~~~~~~~~~~~~~~~~~~~~~~~\times 
        {\rm{tr}}_{g_a,g_c} \left[ \ket{\sqrt{T\eta_o'}\alpha}_{g_a}\bra{\sqrt{T\eta_o'}\alpha} \otimes \ket{-\sqrt{T\eta_o'}\alpha}_{g_c}\bra{-\sqrt{T\eta_o'}\alpha} \right]
        \nonumber
        \\
        &= \frac{\left( 1 - e^{-\eta_oT\alpha^2} \right)^2}{4} 
        \left[ 
        \ket{01}_{a_1b_1} \bra{01}
        \otimes
        \ket{\sqrt{2T}\alpha}_{b_2}\bra{\sqrt{2T}\alpha}
        +
        \ket{10}_{a_1b_1} \bra{10} 
        \otimes
        \ket{-\sqrt{2T}\alpha}_{b_2}\bra{-\sqrt{2T}\alpha}
        \right.
        \nonumber 
        \\
        &~~~~~~~~~~~~~~~~~~~~~~~~ +\left. \left(
        \ket{01}_{a_1b_1} \bra{10} 
        \otimes
        \ket{\sqrt{2T}\alpha}_{b_2}\bra{-\sqrt{2T}\alpha}
        +
        \ket{10}_{a_1b_1} \bra{01} 
        \otimes
        \ket{-\sqrt{2T}\alpha}_{b_2}\bra{\sqrt{2T}\alpha}
        \right) 
        e^{-4(1-T)\alpha^2}
        \right],
    \end{align}
\end{widetext}
where ${\rm{tr}} \Big( \ket{\alpha}\bra{-\alpha} \Big) = e^{-2\alpha^2}$.
Corresponding probability and normalization constant are
\begin{align}
    P_0 = N_0 &= 
    {\rm{tr}}_{\substack{a_1,b_1\\b_2}} \left[ {\rm{tr}}_{\substack{f_a,f_b\\g_a,g_c\\a_2,c}} \Big( \ket{\psi_0}\bra{\psi_0} \Big) \right] 
    \nonumber 
    \\
    &= \frac{\left( 1 - e^{-\eta_oT\alpha^2} \right)^2}{2}.
    \label{append_eq:p0n0}
\end{align}
This leads to the normalized state
\begin{widetext}
    \begin{align}
        &\rho^0_{a_1,b_1,b_2} = \frac{1}{N_0} {\rm{tr}}_{\substack{f_a,f_b\\g_a,g_c\\a_2,c}} \Big( \ket{\psi_0}\bra{\psi_0} \Big) 
        \nonumber 
        \\
        &= \frac{1}{2} 
        \left[ 
        \ket{01}_{a_1b_1} \bra{01}
        \otimes
        \ket{\sqrt{2T}\alpha}_{b_2}\bra{\sqrt{2T}\alpha}
        +
        \ket{10}_{a_1b_1} \bra{10} 
        \otimes
        \ket{-\sqrt{2T}\alpha}_{b_2}\bra{-\sqrt{2T}\alpha}
        \right.
        \nonumber 
        \\
        &~~~~~~~ +\left. \left(
        \ket{01}_{a_1b_1} \bra{10} 
        \otimes
        \ket{\sqrt{2T}\alpha}_{b_2}\bra{-\sqrt{2T}\alpha}
        +
        \ket{10}_{a_1b_1} \bra{01} 
        \otimes
        \ket{-\sqrt{2T}\alpha}_{b_2}\bra{\sqrt{2T}\alpha}
        \right) 
        e^{-4(1-T)\alpha^2}
        \right].
        \label{append_eq:pho0onoff}
    \end{align}
\end{widetext}


\subsection{Final state obtained after the homodyne measurement}
\label{subsec:rhohom}


Charlie now performs the homodyne measurement along the quadrature $X_\theta$ on mode $b_2$.
Here also we consider that that the homodyne instruments are not perfect.
Rather the efficiency of the homodyne detector is given by $\eta_h$.
Similar to the earlier cases here also the imperfect homodyne detector could be modeled as a passive beam splitter with transmittance $\eta_h$.
Now the action of the imperfect homodyne measurement along quadrature $X_\theta$ will lead to the resultant unnormalized state $\rho_{\rm{un}}^{\rm{hom},0}=\bra{X_\theta}_{b_2} {\rm{tr}}_{h_b}\left[ \left(U_{\eta_h}^{h_b,b_2}\right) \rho_{a_1,b_1;b_2}^{0} \otimes \ket{0}_{h_b}\bra{0} \left( U_{\eta_h}^{h_b,b_2} \right)^{\dagger} \right] \ket{X_\theta}_{b_2}$ with normalization $N_0^{\rm{hom}} = {\rm{tr}}_{a_1,b_1} \left( \rho_{\rm{un}}^{\rm{hom},0} \right)$, where 
$X_\theta$ in a mode $a$ is defined as $X_\theta=(ae^{\mathsf{i}\theta} + a^{\dagger}e^{-\mathsf{i}\theta})/2$ and with the eigenvalue equation as $X_\theta$ defined as $X_\theta \ket{X_\theta} = x_\theta \ket{X_\theta}$.

Thus, after the homodyne measurement by Charlie, the residual normalized state between Alice and Bob will be 
\begin{equation}
  \rho_{a_1,b_1} = \frac{\rho_{\rm{un}}^{\rm{hom},0}}{N_0^{\rm{hom}}}.
  \label{append_eq:rhohom}
\end{equation}

In this work we consider the measurement of quadrature operator for the choice of $\theta=\frac{\pi}{2}$, i.e., we consider the {\em momentum-like} quadrature operator $P$.
Now the measurement of $P$ for a coherent state $\ket{\alpha}$ is $\bra{P}\ket{\alpha}= \frac{1}{\pi^{1/4}} e^{-p^2/2} e^{-\alpha^2 - \mathsf{i}\sqrt{2}\alpha p}$.
Now,
\begin{widetext}
    \begin{align}
        &{\rm{tr}}_{h_b} \left[ \left( U_{\eta_h}^{h_b,b_2} \right) \rho_{a_1,b_1;b_2}^0 \otimes \ket{0}_{h_b}\bra{0} \left( U_{\eta_h}^{h_b,b_2} \right)^{\dagger} \right]
        \nonumber 
        \\
        &= \frac{1}{2} {\rm{tr}}_{h_b} \left\lbrace
        \left( U_{\eta_h}^{h_b,b_2} \right)
        \left[
        \ket{01}_{a_1b_1} \bra{01}
        \otimes
        \ket{\sqrt{2T}\alpha}_{b_2}\bra{\sqrt{2T}\alpha}
        +
        \ket{10}_{a_1b_1} \bra{10} 
        \otimes
        \ket{-\sqrt{2T}\alpha}_{b_2}\bra{-\sqrt{2T}\alpha}
        \right.
        \right.
        \nonumber 
        \\
        &~~ +\left.\left. \left(
        \ket{01}_{a_1b_1} \bra{10} 
        \otimes
        \ket{\sqrt{2T}\alpha}_{b_2}\bra{-\sqrt{2T}\alpha}
        +
        \ket{10}_{a_1b_1} \bra{01} 
        \otimes
        \ket{-\sqrt{2T}\alpha}_{b_2}\bra{\sqrt{2T}\alpha}
        \right) 
        e^{-4(1-T)\alpha^2}
        \right]
        \otimes 
        \ket{0}_{h_b}\bra{0}
        \left( U_{\eta_h}^{h_b,b_2} \right)^{\dagger} 
        \right\rbrace
        \nonumber 
        \\
        &= \frac{1}{2} {\rm{tr}}_{h_b} \left[
        \ket{01}_{a_1b_1} \bra{01}
        \otimes
        \ket{\sqrt{2T\eta_h}\alpha}_{b_2}\bra{\sqrt{2T\eta_h}\alpha}
        \otimes 
        \ket{\sqrt{2T\eta_h'}\alpha}_{h_b}\bra{\sqrt{2T\eta_h'}\alpha}
        \right.
        \nonumber 
        \\
        &~~ +\left. \ket{10}_{a_1b_1} \bra{10} 
        \otimes
        \ket{-\sqrt{2T\eta_h}\alpha}_{b_2}\bra{-\sqrt{2T\eta_h}\alpha}
        \otimes
        \ket{-\sqrt{2T\eta_h'}\alpha}_{h_b}\bra{-\sqrt{2T\eta_h'}\alpha}
        \right.
        \nonumber 
        \\
        &~~ +\left(
        \ket{01}_{a_1b_1} \bra{10} 
        \otimes
        \ket{\sqrt{2T\eta_h}\alpha}_{b_2}\bra{-\sqrt{2T\eta_h}\alpha}
        \otimes
        \ket{\sqrt{2T\eta_h'}\alpha}_{h_b}\bra{-\sqrt{2T\eta_h'}\alpha}
        \right.
        \nonumber 
        \\
        &~~ +\left. \left.
        \ket{10}_{a_1b_1} \bra{01} 
        \otimes
        \ket{-\sqrt{2T\eta_h}\alpha}_{b_2}\bra{\sqrt{2T\eta_h}\alpha}
        \otimes
        \ket{-\sqrt{2T\eta_h'}\alpha}_{h_b}\bra{\sqrt{2T\eta_h'}\alpha}
        \right) 
        e^{-4(1-T)\alpha^2} 
        \right]
        \nonumber 
        \\
        &= \frac{1}{2} \left[
        \ket{01}_{a_1b_1} \bra{01}
        \otimes
        \ket{\sqrt{2T\eta_h}\alpha}_{b_2}\bra{\sqrt{2T\eta_h}\alpha}
        + 
        \ket{10}_{a_1b_1} \bra{10} 
        \otimes
        \ket{-\sqrt{2T\eta_h}\alpha}_{b_2}\bra{-\sqrt{2T\eta_h}\alpha}
        \right.
        \nonumber 
        \\
        &~~ +\left.\left(
        \ket{01}_{a_1b_1} \bra{10} 
        \otimes
        \ket{\sqrt{2T\eta_h}\alpha}_{b_2}\bra{-\sqrt{2T\eta_h}\alpha}
        + 
        \ket{10}_{a_1b_1} \bra{01} 
        \otimes
        \ket{-\sqrt{2T\eta_h}\alpha}_{b_2}\bra{\sqrt{2T\eta_h}\alpha}
        \right) 
        e^{-4T(1-\eta_h')\alpha^2}
        e^{-4(1-T)\alpha^2} 
        \right]
        \nonumber 
        \\
        &= \frac{1}{2} \left[
        \ket{01}_{a_1b_1} \bra{01}
        \otimes
        \ket{\sqrt{2T\eta_h}\alpha}_{b_2}\bra{\sqrt{2T\eta_h}\alpha}
        + 
        \ket{10}_{a_1b_1} \bra{10} 
        \otimes
        \ket{-\sqrt{2T\eta_h}\alpha}_{b_2}\bra{-\sqrt{2T\eta_h}\alpha}
        \right.
        \nonumber 
        \\
        &~~ +\left.\left(
        \ket{01}_{a_1b_1} \bra{10} 
        \otimes
        \ket{\sqrt{2T\eta_h}\alpha}_{b_2}\bra{-\sqrt{2T\eta_h}\alpha}
        + 
        \ket{10}_{a_1b_1} \bra{01} 
        \otimes
        \ket{-\sqrt{2T\eta_h}\alpha}_{b_2}\bra{\sqrt{2T\eta_h}\alpha}
        \right) 
        e^{-4(1-T\eta_h)\alpha^2} 
        \right],
    \end{align}
    which leads to 
    \begin{align}
        &\rho_{\rm{un}}^{\rm{hom},0}= \bra{P}_{b_2} {\rm{tr}}_{h_b} \left[ \left( U_{\eta_h}^{h_b,b_2} \right) \rho_{a_1,b_1;b_2}^0 \otimes \ket{0}_{h_b}\bra{0} \left( U_{\eta_h}^{h_b,b_2} \right)^{\dagger} \right] \ket{P}_{b_2}
        \nonumber 
        \\
        &= \frac{1}{2} 
        \bra{P}_{b_2} 
        \left[
        \ket{01}_{a_1b_1} \bra{01}
        \otimes
        \ket{\sqrt{2T\eta_h}\alpha}_{b_2}\bra{\sqrt{2T\eta_h}\alpha}
        + 
        \ket{10}_{a_1b_1} \bra{10} 
        \otimes
        \ket{-\sqrt{2T\eta_h}\alpha}_{b_2}\bra{-\sqrt{2T\eta_h}\alpha}
        \right.
        \nonumber 
        \\
        &~~ +\left.\left(
        \ket{01}_{a_1b_1} \bra{10} 
        \otimes
        \ket{\sqrt{2T\eta_h}\alpha}_{b_2}\bra{-\sqrt{2T\eta_h}\alpha}
        + 
        \ket{10}_{a_1b_1} \bra{01} 
        \otimes
        \ket{-\sqrt{2T\eta_h}\alpha}_{b_2}\bra{\sqrt{2T\eta_h}\alpha}
        \right) 
        e^{-4(1-T\eta_h)\alpha^2} 
        \right] 
        \ket{P}_{b_2}
        \nonumber 
        \\
        &= \frac{e^{-p^2}}{2\sqrt{\pi}}
        \left[
        \ket{01}_{a_1b_1} \bra{01}
        e^{-4T\eta_h\alpha^2}
        + 
        \ket{10}_{a_1b_1} \bra{10} 
        e^{-4T\eta_h\alpha^2}
        \right.
        \nonumber 
        \\
        &~~ +\left.\left(
        \ket{01}_{a_1b_1} \bra{10} 
        e^{-4T\eta_h\alpha^2 - 4\mathsf{i}\sqrt{T\eta_h}\alpha p}
        + 
        \ket{10}_{a_1b_1} \bra{01} 
        e^{-4T\eta_h\alpha^2 + 4\mathsf{i}\sqrt{T\eta_h}\alpha p}
        \right) 
        e^{-4(1-T\eta_h)\alpha^2} 
        \right]
        \nonumber 
        \\
        &= \frac{e^{-p^2} e^{-4T\eta_h\alpha^2}}{2\sqrt{\pi}}
        \left[
        \ket{01}_{a_1b_1} \bra{01}
        + 
        \ket{10}_{a_1b_1} \bra{10} 
        +
        \left(
        \ket{01}_{a_1b_1} \bra{10} 
        e^{-4\mathsf{i}\sqrt{T\eta_h}\alpha p}
        +
        \ket{10}_{a_1b_1} \bra{01} 
        e^{4\mathsf{i}\sqrt{T\eta_h}\alpha p}
        \right) 
        e^{-4(1-T\eta_h)\alpha^2} 
        \right],
    \end{align}
    with $N_0^{\rm{hom}} = {\rm{tr}}_{a_1,b_1}\left( \rho_{\rm{un}}^{\rm{hom},0} \right) = \frac{e^{-p^2} e^{-4T\eta_h\alpha^2}}{\sqrt{\pi}}$. 
    
    Therefore, 
    \begin{align}
        \rho_{a_1,b_1} &= \frac{\rho_{\rm{un}}^{\rm{hom},0}}{N_0^{\rm{hom}}}
        \nonumber 
        \\
        &= \frac{1}{2} \left[ 
        \ket{01}_{a_1b_1} \bra{01} + \ket{10}_{a_1b_1} \bra{10} + \left(
        \ket{01}_{a_1b_1}\bra{10} e^{-4\mathsf{i}\sqrt{T\eta_h}\alpha p}
        + \ket{10}_{a_1b_1}\bra{01} e^{4\mathsf{i}\sqrt{T\eta_h}\alpha p}
        \right) e^{-4(1-T\eta_h)\alpha^2} 
        \right]
        \nonumber 
        \\
        &= \frac{1}{2} \args{ \ket{01}_{a_1b_1}\bra{01} + \ket{10}_{a_1b_1} \bra{10} + h\argp{ g\ket{01}_{a_1b_1}\bra{10} + g^*\ket{10}_{a_1b_1}\bra{01} }}
        \label{suppl_eq:final_entangled_state},
    \end{align}
    where $h = e^{-4(1-T\eta_h)\alpha^2}$ and $g = e^{-4\mathsf{i}\sqrt{T\eta_h}\alpha p}$.
    The probability of obtaining this final state is given by \eqref{append_eq:p0n0} $P_0 = \frac{\left( 1 - e^{-\eta_oT\alpha^2} \right)^2}{2}$.
\end{widetext}

\section{Logarithmic negativity of hybrid entangled states undergoing photon loss}
\label{append_sec:logneg_hybrid_photonloss}

In our protocol we use the CV part of the HE state for transmission via a lossy quantum channel. 
It can be shown that under photon losses in the CV part, an HE state can still retain correlations for a particular value of $\alpha$.

We analyse the amount of correlations that a HE state retains after its CV system is transmitted via a lossy quantum channel. Upon transmission the CV part undergoes photon loss which is directly dependent on the value of $\alpha$ chosen. We show that the correlations in a HE state after its CV part has undergone transmission loss is a non-monotonic function of its coherent amplitude. Specifically, we evaluate the logarithmic negativity~\cite{ZHP98, P05} of the initial HE state as a function of transmission loss. We find that for $\alpha \approx 0.5$, the HE state is highly robust against noise.

Let us consider the HE state
\begin{equation}
   \ket{\psi}_{ab} = \frac{1}{\sqrt{2}} \Big( \ket{0}_a\ket{\alpha}_b + \ket{1}_a\ket{-\alpha}_b \Big).
\end{equation}
Suppose that the mode $b$ undergoes photon loss.
The process of photon loss can be equivalently modeled as passage through a beam splitter with reflectivity $R$ ($0\leq R\leq 1$) while the other input to the beam splitter is taken to be vacuum. 
In such a case, the beam splitter matrix is
$
\begin{pmatrix} 
\sqrt{1-R} & \sqrt{R} \\
-\sqrt{R} & \sqrt{1-R}
\end{pmatrix},
$
where $R=0$ and $R=1$ stand for zero photon loss and complete photon loss, respectively. 
To that end, let us consider that the mode $b$ passes through such a beam splitter while the other input is at $\ket{0}$ in mode $c$.
As a consequence, the total state after passage through the beam splitter becomes
\begin{widetext}
    \begin{equation}
        \ket{\psi}_{ab} \otimes \ket{0}_c \xrightarrow{\rm{bs}} \frac{1}{\sqrt{2}} \left( \ket{0}_a \ket{\sqrt{1-R}\alpha}_b \ket{\sqrt{R}\alpha}_c +  \ket{1}_a \ket{-\sqrt{1-R}\alpha}_b \ket{-\sqrt{R}\alpha}_c \right).
    \end{equation}

    Subsequently, the two-mode state in modes $a$ and $b$ after photon loss is obtained by tracing over the ancillary mode $c$ as
    \begin{align}
        \rho_{ab}^{\rm{loss}} &= {\rm{tr}}_{c} \left[ \frac{1}{2} \left( 
        \ket{0,\sqrt{1-R}\alpha}_{ab} \bra{0,\sqrt{1-R}\alpha} \otimes \ket{\sqrt{R}\alpha}_c \bra{\sqrt{R}\alpha} 
        +
        \ket{1,-\sqrt{1-R}\alpha}_{ab} \bra{1,-\sqrt{1-R}\alpha} \otimes \ket{-\sqrt{R}\alpha}_c \bra{-\sqrt{R}\alpha}
        \right.
        \right.
        \nonumber 
        \\
        &~~\left.\left.
        + \ket{0,\sqrt{1-R}\alpha}_{ab} \bra{1,-\sqrt{1-R}\alpha} \otimes \ket{\sqrt{R}\alpha}_c \bra{-\sqrt{R}\alpha} 
        +
        \ket{1,-\sqrt{1-R}\alpha}_{ab} \bra{0,\sqrt{1-R}\alpha} \otimes \ket{-\sqrt{R}\alpha}_c \bra{\sqrt{R}\alpha}
        \right) \right]
        \nonumber 
        \\
        &= \frac{1}{2} \left( \ket{0,\sqrt{1-R}\alpha}_{ab} \bra{0,\sqrt{1-R}\alpha} 
        +
        \ket{1,-\sqrt{1-R}\alpha}_{ab} \bra{1,-\sqrt{1-R}\alpha}
        +
        e^{-2R\alpha^2} \ket{0,\sqrt{1-R}\alpha}_{ab} \bra{1,-\sqrt{1-R}\alpha}
        \right. 
        \nonumber
        \\
        &~~\left.
        +
        e^{-2R\alpha^2} \ket{1,-\sqrt{1-R}\alpha}_{ab} \bra{0,\sqrt{1-R}\alpha}
        \right)
        \nonumber 
        \\
        &= \frac{1}{2} \left[ \ket{0}_a\bra{0} \otimes \ket{\sqrt{1-R}\alpha}_b\bra{\sqrt{1-R}\alpha} +
        \ket{1}_a\bra{1} \otimes \ket{-\sqrt{1-R}\alpha}_b\bra{-\sqrt{1-R}\alpha}
        \right.
        \nonumber 
        \\
        &~~\left. 
        + e^{-2R\alpha^2} \left( 
        \ket{0}_a\bra{1} \otimes \ket{\sqrt{1-R}\alpha}_b\bra{-\sqrt{1-R}\alpha} 
        + 
        \ket{1}_a\bra{0} \otimes \ket{-\sqrt{1-R}\alpha}_b\bra{\sqrt{1-R}\alpha}
        \right)\right].
        \label{append_eq:hes_loss}
    \end{align}
\end{widetext}

In order to evaluate the entanglement content in this state, we use logarithmic negativity as a measure of entanglement. For a bipartite state $\rho_{ab}^{\rm{loss}}$ it is defined as 
$E_{N} \left( \rho_{ab}^{\rm{loss}} \right) = \log_2 \left|\left| \left( \rho_{ab}^{\rm{loss}} \right)^{\rm{P.T.}} \right|\right|_1$, 
where $\left|\left| . \right|\right|_1$ is the trace norm and P.T. stands for partial transpose over any one of the modes $a$ or $b$.
We evaluate the logarithmic negativity for the state after photon loss in Eq.~\eqref{append_eq:hes_loss} numerically and is shown in the main text.

\section{Logarithmic negativity of the shared entangled states}
\label{append_sec:logneg_finalstate}

In order to evaluate the entanglement content in this state, we use logarithmic negativity as a measure of entanglement. For a bipartite state $\rho_{a_1b_1}$ it is defined as 
$E_{N} \left( \rho_{a_1b_1} \right) = \log_2 \left|\left|  \rho_{a_1b_1}^{\rm{P.T.}} \right|\right|_1$, 
where $\left|\left| . \right|\right|_1$ is the trace norm and P.T. stands for partial transpose over any one of the modes $a_1$ or $b_1$.
Here, we evaluate the logarithmic negativity for the shared entangled state between Alice and Bob~\eqref{suppl_eq:final_entangled_state} under partial transposition over mode $b_1$. The resultant state after the partial transpose is written as
\begin{align}
    \rho_{a_1,b_1}^\text{P.T.} &= \frac{1}{2} \left[
    \ket{01}_{a_1b_1} \bra{01} + \ket{10}_{a_1b_1} \bra{10} 
    \right.
    \nonumber 
    \\
    &~~\left.
    + h\left( g\ket{00}_{a_1b_1} \bra{11} + g^*\ket{11}_{a_1b_1} \bra{00} 
    \right) 
    \right],
\end{align}
where $h=e^{-4(1-T\eta_h)\alpha^2}$ and $g=e^{-4\mathsf{i}\sqrt{T\eta_h}\alpha p}$. 
This leads to the eigenvalues of $\rho_{a_1,b_1}^\text{P.T.}$ as $\lambda_1 = \lambda_2 = \frac{1}{2}$, $\lambda_3 = \frac{h}{2}$ and $\lambda_4 = -\frac{h}{2}$.
As a consequence, the logarithmic negativity of $\rho_{a_1b_1}$ is given as
\begin{equation}
E_N\left( \rho_{a_1b_1} \right) = \log_2\left( \sum_{k=1}^4 |\lambda_k| \right) = \log_2\left( 1 + h \right).
\end{equation}

However, it should be noted that the final state, $\rho_{a_1b_1}$, is only produced with a probability $P_0$. 
Consequently, the entanglement between the parties is effectively given as $E_n\argp{\rho_{a_1b_1}} = P_0 \log_2\left( 1 + h \right)$. 
The reason for multiplying with the probability $P_0$ is because it determines the rate of generation of the resultant entangled state. 
As an example consider $\alpha = 0$, the HE state \eqref{suppl_eq:hes_def} is effectively a separable state and therefore cannot yield any correlations after swapping. 
This behavior is captured by the fact that the final entangled state is produced with probability $0$. 
However, if we only look at the state $\rho_{a_1b_1}$, we find that it is maximally entangled with logarithmic negativity equal to $1$. 
Therefore, it is necessary to include the rate of production in the analysis of entanglement of the final state.

\section{Analysis of secure key rate}
\label{append_sec:calculations_keyrate}
In this section we first provide a description of the optimal strategy of an eavesdropper Eve, namely an entangling cloner attack. Next, provide a detailed analysis of the secure key rate under this strategy by Eve. 

\subsection{Evaluating the secured key rate}
In order to evaluate the secure key rate, we assume the existence of an eavesdropper Eve with system $E$. 
We assume that Eve can potentially collaborate with the untrusted party Charlie while also having access to the two quantum channels which are used to transmit the CV systems. 
We also consider that Eve can perfrom an entangling cloner attack on each of two the quantum channels~\cite{GCW03, OSB15,POS15}. 
However, the most general attack strategy with Eve is a two-mode correlated attack (one mode for each quantum channel). 
Since, Alice and Bob use a CV system for transmission purposes, the aforementioned attacks have been shown to be the optimal choices in such a case. Moreover, since the quantum channels are assumed to be non-interacting and spatially well separated, the two-mode correlated attack reduces to two independent single-mode entangling cloner attacks.

Specifically, a single mode entangling cloner attack assumes that Eve can split the incoming CV states in both the channels independently using a BS with transmittance $T$ which equal to the loss of the Alice-Charlie and Bob-Charlie channels (assuming that the loss in both the channels is same). 
The two input modes for this BS correspond to the quantum state being transmitted and a vacuum state (or a thermal state if we consider thermal noise in the channels). 
Eve, then stores the reflected states in a quantum memory while the transmitted states are sent to Charlie via identity channels having no loss. 
Subsequently, Eve can then perform a joint measurement on the two retained states (corresponding to Alice-Charlie and Bob-Charlie channels) which are stored in a quantum memory and try to guess the key of Alice or Bob based on the outcomes observed. 
However, Alice and Bob can estimate the transmission losses of their respective channels given by $T$. 
As a consequence, the maximum information that can obtained by Eve becomes a function of the channel loss parameter $T$ and the publicly declared results by Charlie which in turn is bounded by the Holevo bound $\chi(A:E)$~\cite{POS15, OSB15}.

Since, Alice and Bob share the state $\rho_{a_1b_1}$ with probability for a detailed derivation) $P_0$, the secure key rate $r$ between Alice and Bob is then given as
\begin{equation}
    r \geq P_0 \left[I(A:B) - \chi(A:E)\right] ~~\rm{s.t.}~~ P_0=\frac{\left( 1 - e^{-\eta_oT\alpha^2} \right)^2}{2}.
    \label{suppl_eq:secure_key}
\end{equation}

Evaluating the mutual information between Alice and Bob is relatively simple and is accomplished by using their observed joint statistics. 
If Alice and Bob perform a measurement corresponding to observables $A$ and $B$, the mutual information between the two parties sharing a state $\rho_{ab}$ is given as $I(A : B) = H(A) + H(B) - H(A, B)$, 
where $H(A)$ (and $H(B)$) is the Shannon entropy corresponding to the observable $A$ (and $B$) measured on the state $\rho_a = \text{tr}_b\left(\rho_{ab}\right)$ and, $H(A, B)$ is the Shannon entropy of the observables jointly measured on the state $\rho_{ab}$.

\subsection{Calculation of $I(A:B)$ and $\chi(A:E)$}
In order to evaluate the mutual information we first look at the final state that is shared between Alice and Bob which is given as
\begin{align}
    \rho_{a_1b_1} &= \frac{1}{2} \left[ \ket{01}_{a_1b_1} \bra{01} + \ket{10}_{a_1b_1} \bra{10} 
    \right.
    \nonumber 
    \\
    &\left. 
    + h\left( g\ket{01}_{a_1b_1} \bra{10} + g^*\ket{10}_{a_1b_1} \bra{01} \right) 
    \right]
\label{suppl_eq:final_state_qkd},
\end{align}
where $h = e^{-4(1-T\eta_h)\alpha^2}$ and $g = e^{-4\mathsf{i}\sqrt{T\eta_h}\alpha p}$ with the reduced states of Alice and Bob as
\begin{equation}
\rho_{a_1} = \rho_{b_1} = \frac{1}{2} \argp{ \ket 0\bra 0 + \ket 1\bra 1}.
\end{equation}

In the QKD protocol we consider that both Alice and Bob choose the observable $M = \sigma_Z$ to generate a key. The corresponding projective measurement can then be written as $\lbrace \Pi_0, \Pi_1\rbrace$, where $\Pi_0 = \ket 0\bra 0$ and $\Pi_1 = \mathds{1} - \Pi_0 = \ket 1\bra 1$.
We also consider that the photon number detectors are imperfect having efficiency $\eta_d$.
A general $m$-photon detector with efficiency $\eta_d$ is described by the measurement operators
\begin{equation}
\Pi_m(\eta_d) = \eta_d^m \sum_k \argp{1-\eta_d}^k \ket{k+m}\bra{k+m}.
\end{equation}

In view of the fact that in our scheme we have only two outcomes (corresponding to $\Pi_0$ and $\mathds{1} - \Pi_0$), an imperfect measurement of $\sigma_Z$ then corresponds to measurement operators
\begin{subequations}
\begin{align}
& \Pi_0(\eta_d) = \ket 0\bra 0 + (1-\eta_d)\ket 1\bra 1,\\
&\Pi_1(\eta_d) = \mathds{1} - \Pi_0 = \eta_d \ket 1\bra 1.
\end{align}
\end{subequations}

We first consider the imperfect measurement of $\sigma_Z$ on Alice's reduced state $\rho_{a_1}$. The outcome $\Pi_0(\eta_d)$ occurs with probability $p_0 = \frac{1}{2}\args{1 + (1-\eta_d)} = \frac{2-\eta_d}{2}$ while the outcome $\Pi_1(\eta_d)$ occurs with probability $p_1 = 1 - p_0 = \frac{\eta_d}{2}$.
As a consequence, the Shannon entropy of imperfectly measuring $\sigma_Z$ on Alice's reduced state is given as
\begin{align}
    H(\sigma_3) &= -p_0\log_2 p_0 - p_1\log_2 p_1 
    \nonumber 
    \\
    &= -\argp{\frac{2-\eta_d}{2}}\log_2 \argp{\frac{2-\eta_d}{2}} - \frac{\eta_d}{2}\log_2 \frac{\eta_d}{2}.
    \nonumber 
    \\
    &=1 - \frac{2-\eta_d}{2}\log_2(2-\eta_d) - \frac{\eta_d}{2}\log_2\eta_d 
    \label{suppl_eq:shannon_entropy_partial}
\end{align}

Similarly, it can be seen that the same expression also holds true for the imperfect measurement of $\sigma_Z$ on Bob's reduced state.
The measurement operators corresponding to the case when Alice and Bob jointly (and imperfectly) measure $\sigma_Z$ on their respective reduced states are 
\begin{subequations}
    \begin{align}
    \Pi_{00}(\eta_d) &= \ket{00}\bra{00} + (1-\eta_d)^2\ket{11}\bra{11} + (1-\eta_d)\ket{01}\bra{01} 
    \nonumber 
    \\
    &~~ + (1-\eta_d)\ket{10}\bra{10},
    \\
    \Pi_{01}(\eta_d) &= \eta_d\ket{01}\bra{01} + \eta_d(1-\eta_d)\ket{11}\bra{11},
    \\
    \Pi_{10}(\eta_d) &= \eta_d\ket{10}\bra{10} + \eta_d(1-\eta_d)\ket{11}\bra{11},
    \\
    \Pi_{11}(\eta_d) &= \eta_d^2\ket{11}\bra{11},
    \end{align}
\end{subequations}
which occur with probabilities $p_{00} = 1-\eta_d$, $p_{01} = \frac{\eta_d}{2}$, $p_{10} = \frac{\eta_d}{2}$, and $p_{11} = 0$, respectively.
As a consequence, the Shannon entropy for the joint measurement becomes
\begin{widetext}
    \begin{align}
        H(\sigma_Z, \sigma_Z) &= - p_{00}\log_2 p_{00} - p_{01}\log_2 p_{01} - p_{10}\log_2 p_{10} - p_{11}\log_2 p_{11} 
        \nonumber 
        \\
        &= -(1-\eta_d)\log_2 (1-\eta_d) - \eta_d\log_2\frac{\eta_d}{2} 
        = \eta_d - (1-\eta_d)\log_2(1-\eta_d) - \eta_d\log_2\eta_d.
    \end{align}
\end{widetext}

The mutual information between Alice and Bob cab then be written as
\begin{align}
    &I(A:B) = H(\sigma_Z) + H(\sigma_Z) - H(\sigma_Z, \sigma_Z)
    \nonumber 
    \\
    &= (2-\eta_d) - \big[(2-\eta_d)\log_2(2-\eta_d) - (1-\eta_d)\log_2(1-\eta_d)\big]
    \label{suppl_eq:mutual_inf_sigma3}.
\end{align}

Evidently, in absence of any imperfection ($\eta_d=1$) one obtains perfect correlation, i.e, $\lim_{\eta_d\rightarrow 1}I(A:B) = 1$.
Next, we evaluate the Holevo bound $\chi(A:E)$.

In order to evaluate the Holevo bound, we assume that Eve has access to a purification of the state $\rho_{a_1b_1}$, which we denote by $\rho_{a_1b_1E}$, such that $\rho_e = \text{tr}_{a_1b_1}(\rho_{a_1b_1e})$ is the reduced state of Eve. 
Moreover, we also assume that Alice's measurement outcomes are represented by rank-1 operators.
Since $\rho_{a_1b_1e}$ is pure by definition, we have $S(\rho_{a_1b_1}) = S(\rho_e)$, where $S(X)$ is the Von Neumann entropy of a system $X$. 
Moreover, if Alice's measurement outcomes are represented by rank-1 operators, then the reduced state of Bob and Eve conditioned on Alice's outcome $x$, given by $\rho_{b_1e|x}$, is also pure. 
Therefore, by definition of Von Neumann entropy, we have $S(\rho_{e|x}) = S(\rho_{b_1|x})$, where $\rho_{b_1|x}$ is the reduced state of Bob conditioned on Alice's outcome $x$.
In this case, the Holevo bound can then be written as~\cite{BP12}
\begin{equation}
\chi(A:E) = S(\rho_{a_1b_1}) - \sum_x p_x S\left( \rho_{b_1|x} \right).
\label{suppl_eq:def_holevo_bound}
\end{equation}

For the state $\rho_{a_1b_1}$ as given in Eq.~\eqref{suppl_eq:final_state_qkd}, its eigenvalues are given as $\lambda_\pm = \frac{1\pm h}{2}$ leading to the von-Neumann entropy as
\begin{align}
    &S(\rho_{a_1b_1}) = -\lambda_+\log_2 \lambda_+ - \lambda_-\log_2 \lambda_- 
    \nonumber 
    \\
    &= -\argp{\frac{1 + h}{2}}\log_2 \argp{\frac{1 + h}{2}} - \argp{\frac{1 - h}{2}}\log_2 \argp{\frac{1 - h}{2}}
    \nonumber 
    \\
    &= 1 - \frac{1}{2}\left[(1 + h)\log_2(1 + h) + (1 - h)\log_2(1 - h) \right].
\end{align}

On the other hand, the reduced states of Bob corresponding to the two outcomes of the imperfect measurement of $\sigma_Z$ by Alice are given as
\begin{subequations}
    \begin{align}
        \rho_{b_1|0} &= \frac{\text{tr}_{a_1}\args{\rho_{a_1b_1}\Pi_{0}(\eta_d)}}{\text{tr}\args{\rho_{a_1b_1}\Pi_{0}(\eta_d)}}
        \nonumber 
        \\
        &= \frac{1}{2-\eta_d}\argp{\ket 1\bra 1 + (1-\eta_d)\ket 0\bra 0},
        \\
        \rho_{b_1|1} &= \frac{\text{tr}_{a_1}\args{\rho_{a_1b_1}\Pi_{1}(\eta_d)}}{\text{tr}\args{\rho_{a_1b_1}\Pi_{1}(\eta_d)}}
        = \ket 1\bra 1,
    \end{align}
\end{subequations}
that lead to 
\begin{align}
    &\sum_{x=0}^{1} p_x S\left( \rho_{b_1|x} \right) = \frac{2-\eta_d}{2}\left[ -\argp{\frac{1}{2-\eta_d}}\log_2 \argp{\frac{1}{2-\eta_d}} 
    \right.
    \nonumber 
    \\
    &~~\left. 
    - \argp{\frac{1-\eta_d}{2-\eta_d}}\log_2 \argp{\frac{1-\eta_d}{2-\eta_d}} 
    \right]
    \nonumber 
    \\
    &= \frac{1}{2} \big[ (2-\eta_d)\log_2(2-\eta_d) - (1-\eta_d)\log_2(1-\eta_d) \big].
\end{align}

As a consequence, the Holevo bound can be written as
    \begin{align}
    &\chi(A:E) = S(\rho_{a_1b_1}) - \sum_x p_x S\left( \rho_{b_1|x} \right)
    \nonumber 
    \\
    &= 1 - \frac{1}{2}\left[ (1+h)\log_2(1+h) + (1-h)\log_2(1-h) \right] 
    \nonumber 
    \\
    &~~ - \frac{1}{2} \left[ (2-\eta_d)\log_2(2-\eta_d) - (1-\eta_d)\log_2(1-\eta_d) \right]
    \label{suppl_eq:holevo_bound_sigma3}.
\end{align}

By plugging the mutual information given in Eq.~\eqref{suppl_eq:mutual_inf_sigma3} and the Holevo bound, given in Eq.~\eqref{suppl_eq:holevo_bound_sigma3}, in the secure key rate, given by Eq.~\eqref{suppl_eq:secure_key}, we obtain
\begin{align}
    r &\geq p_0 \args{I(A:B) - \chi(A:E)}
    \nonumber 
    \\
    &= p_0 \left\lbrace 1 -\eta_d + \frac{1}{2}\left[ (1+h)\log_2(1+h) + (1-h)\log_2(1-h) \right] 
    \right.
    \nonumber 
    \\
    &~~\left. 
    - \frac{1}{2} \left[ (2-\eta_d)\log_2(2-\eta_d) - (1-\eta_d)\log_2(1-\eta_d) \right] 
    \right\rbrace
    \label{suppl_eq:final_keyrate}.
\end{align}

\section{Fidelity between the HE states passed through loss-only and lossy$+$noise channels}
\label{append_sec:hes_fid_loss_noise}

In this section, we analyze the impact of thermal noise present in the quantum channel between Alice (Bob) and Charlie. We show that the final state after a loss-only channel is almost equivalent to the final state after passing through a channel characterized by loss and thermal noise. Using this result, we aim to reduce the complexity of the calculation by only focusing on loss-only quantum channels connecting all the parties.

\subsection{General state after channel transmission}
\label{subsec:genstate_channel}
Let us consider that an incoming signal passes through a lossy channel having transmittance $T$ with additional thermal noise. 
Mathematically, the channel transmission can be written as $U_{\rm{ch}}:\rho_{\rm{s,in}}\rightarrow \rho_{\rm{s,out}}$, where $\rho_{\rm{s,in}}$ ($\rho_{\rm{s,in}}$) is input (output) state of the channel.
Such transmission can be modeled as follows.
First, the incoming state (in mode $a$) is mixed with an ancilla initialized in a thermal state (in mode $b$) via a beam splitter (BS) with transmittance $T$ and two output modes.
Subsequently, output of the quantum channel is obtained by tracing out the outgoing ancilla mode of the BS.

The action of a BS with transmittance $T$ is described in terms of a unitary operation $U^{ab}_T$ on the input modes $a$ and $b$ that leads to the transformation matrix between the input and the output modes, labelled by $a'$ and $b'$, as
\begin{equation}
	\begin{pmatrix} 
\hat{a} \\
\hat{b}
	\end{pmatrix}\rightarrow
	\begin{pmatrix} 
\hat{a}' \\
\hat{b}'
	\end{pmatrix} = 
	\begin{pmatrix} 
\sqrt{T} & \sqrt{1-T} \\
-\sqrt{1-T} & \sqrt{T}
	\end{pmatrix}
	\begin{pmatrix} 
\hat{a} \\
\hat{b}
	\end{pmatrix},
\end{equation}
where $\hat{a}$ corresponds to the annihilation operator for the mode $a$ and
$T = 0.5$ represents a balanced ($50:50$) BS.
As a consequence, the action of the channel on a coherent state ($\ket\alpha$) in mode $a$ is described as $U^{ab}_T\ket\alpha_{a}\otimes \ket \beta_b\rightarrow \ket\alpha_{a^{'}}\otimes \ket \beta_{b^{'}} = \ket{\sqrt{T}\alpha + \sqrt{1-T}\beta}_a\otimes \ket{\sqrt{T}\beta - \sqrt{1-T}\alpha}_b = \ket{\sqrt{T}\alpha + \sqrt{1-T}\beta, \sqrt{T}\beta - \sqrt{1-T}\alpha}_{ab}$.

Let us consider that, in the coherent state basis, the incoming signal is in the quantum state given by $\rho_{\rm{s,in}} = \int \frac{d^2\alpha}{\pi}\frac{d^2\beta}{\pi} C(\alpha,\beta) \ket\alpha_a\bra\beta$ while the ancilla thermal state described as $\rho_{\rm{anc,th}} = (1-x)\sum_k x^k \ket k_b\bra k = (1-x)\int \frac{d^2\eta}{\pi}\frac{d^2\zeta}{\pi} e^{-\frac{|\eta|^2+|\zeta|^2}{2}+x\eta^*\zeta} \ket\eta_b\bra\zeta$, where $x=\frac{\bar{n}}{1+\bar{n}}$ such that $\bar{n}$ is the average number of thermal photon.
In view of the action of a BS on the coherent states, one can easily show that
\begin{widetext}
\begin{align}
\rho_{\rm{out}} &= U^{ab}_T \rho_{\rm{s,in}} \otimes \rho_{\rm{anc,th}} \argp{U^{ab}_T} ^{\dagger}
\nonumber
\\
&= (1-x)\int \frac{d^2\alpha}{\pi}\frac{d^2\beta}{\pi} \frac{d^2\eta}{\pi}\frac{d^2\zeta}{\pi} 
C(\alpha,\beta) e^{-\frac{|\eta|^2+|\zeta|^2}{2}+x\eta^*\zeta} 
\args{U^{ab}_T \ket{\alpha,\eta}_{ab} \bra{\beta,\zeta} \argp{U^{ab}_T} ^{\dagger}}
\nonumber
\\
&= (1-x)\int \frac{d^2\alpha}{\pi}\frac{d^2\beta}{\pi} \frac{d^2\eta}{\pi}\frac{d^2\zeta}{\pi} 
C(\alpha,\beta) e^{-\frac{|\eta|^2+|\zeta|^2}{2}+x\eta^*\zeta}
\ket{\sqrt{T}\alpha + \sqrt{1-T}\eta,\sqrt{T}\eta - \sqrt{1-T}\alpha}
\nonumber 
\\
&~~~~\otimes
\bra{\sqrt{T}\beta + \sqrt{1-T}\zeta,\sqrt{T}\zeta - \sqrt{1-T}\beta}.
\end{align}
	
Consequently, the channel output signal state becomes
\begin{align}
&U_{\rm{ch}}:\rho_{\rm{s,in}} \rightarrow \rho_{\rm{s,out}} = \rm{Tr}_{\rm{anc}} \argp{\rho_{\rm{out}}}
\nonumber
\\
&= (1-x)\int \frac{d^2\alpha}{\pi}\frac{d^2\beta}{\pi} \frac{d^2\eta}{\pi}\frac{d^2\zeta}{\pi} 
C(\alpha,\beta) e^{-\frac{|\eta|^2+|\zeta|^2}{2}+x\eta^*\zeta}
\ket{\sqrt{T}\alpha + \sqrt{1-T}\eta}_{a}
\bra{\sqrt{T}\beta + \sqrt{1-T}\zeta}
\nonumber 
\\
&~~~~\times
\left\langle \sqrt{T}\zeta - \sqrt{1-T}\beta \ket{\sqrt{T}\eta - \sqrt{1-T}\alpha} \right.
\nonumber
\\
&= (1-x)\int \frac{d^2\alpha}{\pi}\frac{d^2\beta}{\pi} \frac{d^2\eta}{\pi}\frac{d^2\zeta}{\pi}  
C(\alpha,\beta) e^{-\frac{1-T}{2}\argp{|\alpha|^2 + |\beta|^2} + (1-T)\alpha\beta^*}
e^{-\frac{1+T}{2}\argp{|\eta|^2+|\zeta|^2} + x\eta^*\zeta + T\eta\zeta^*}
\nonumber
\\
&~~~~\times e^{\sqrt{T(1-T)}\args{\frac{\alpha^*\eta + \alpha\eta^*}{2} + \frac{\beta^*\zeta + \beta\zeta^*}{2} - \argp{\beta^*\eta + \alpha\zeta^*}}}
\ket{\sqrt{T}\alpha + \sqrt{1-T}\eta}_{a} \bra{\sqrt{T}\beta + \sqrt{1-T}\zeta}
\nonumber
\\
&= (1-x)\int \frac{d^2\alpha}{\pi}\frac{d^2\beta}{\pi} \frac{d^2\eta}{\pi}\frac{d^2\zeta}{\pi}  
C(\alpha,\beta) e^{-\frac{1-T}{2}\argp{|\alpha|^2 + |\beta|^2} + (1-T)\alpha\beta^*}
e^{-\frac{1+T}{2}\argp{|\eta|^2+|\zeta|^2} + x\eta^*\zeta + T\eta\zeta^*}
\nonumber
\\
&~~~~\times e^{\sqrt{T(1-T)}\args{\frac{\alpha^*\eta + \alpha\eta^*}{2} + \frac{\beta^*\zeta + \beta\zeta^*}{2} - \argp{\beta^*\eta + \alpha\zeta^*}}}
\int \frac{d^2\lambda}{\pi}\frac{d^2\omega}{\pi} \ket\lambda\bra\omega
\left\langle\lambda \ket{\sqrt{T}\alpha + \sqrt{1-T}\eta} \bra{\sqrt{T}\beta + \sqrt{1-T}\zeta} \omega\right\rangle 
\nonumber
\\
&= (1-x)\int \frac{d^2\alpha}{\pi}\frac{d^2\beta}{\pi} \frac{d^2\eta}{\pi}\frac{d^2\zeta}{\pi}  
C(\alpha,\beta) e^{-\frac{1-T}{2}\argp{|\alpha|^2 + |\beta|^2} + (1-T)\alpha\beta^*}
e^{-\frac{1+T}{2}\argp{|\eta|^2+|\zeta|^2} + x\eta^*\zeta + T\eta\zeta^*}
\nonumber
\\
&~~~~\times e^{\sqrt{T(1-T)}\args{\frac{\alpha^*\eta + \alpha\eta^*}{2} + \frac{\beta^*\zeta + \beta\zeta^*}{2} - \argp{\beta^*\eta + \alpha\zeta^*}}}
\int \frac{d^2\lambda}{\pi}\frac{d^2\omega}{\pi} \ket\lambda\bra\omega
e^{-\frac{|\lambda|^2 + |\sqrt{T}\alpha + \sqrt{1-T}\eta|^2}{2} + \lambda^*\argp{\sqrt{T}\alpha + \sqrt{1-T}\eta}}
\nonumber
\\
&~~~~\times 
e^{-\frac{|\omega|^2 + |\sqrt{T}\beta + \sqrt{1-T}\zeta|^2}{2} + \argp{\sqrt{T}\beta^* + \sqrt{1-T}\zeta^*}\omega}
\nonumber
\\
&= (1-x)\int \frac{d^2\lambda}{\pi}\frac{d^2\omega}{\pi} 
e^{-\frac{|\lambda|^2 + |\omega|^2}{2}} \ket\lambda\bra\omega
\int \frac{d^2\alpha}{\pi}\frac{d^2\beta}{\pi} C(\alpha,\beta) e^{-\frac{|\alpha|^2 + |\beta|^2}{2} + (1-T)\alpha\beta^* + \sqrt{T}(\lambda^*\alpha + \omega\beta^*)}
\nonumber
\\
&~~~~\times
\int \frac{d^2\eta}{\pi}\frac{d^2\zeta}{\pi} e^{-\argp{|\eta|^2 + |\zeta|^2} + x\eta^*\zeta + T\eta\zeta^*} 
e^{-\sqrt{T(1-T)}\argp{\beta^*\eta + \alpha\zeta^*} + \sqrt{1-T}\argp{\lambda^*\eta + \omega\zeta^*}}
\nonumber
\\
&= \frac{1-x}{1-Tx}\int \frac{d^2\lambda}{\pi}\frac{d^2\omega}{\pi} 
e^{-\frac{|\lambda|^2 + |\omega|^2}{2} + \frac{x(1-T)}{1-Tx}\lambda^*\omega} \ket\lambda\bra\omega
\int \frac{d^2\alpha}{\pi}\frac{d^2\beta}{\pi} C(\alpha,\beta) e^{-\frac{|\alpha|^2 + |\beta|^2}{2} + \frac{1-T}{1-Tx}\alpha\beta^* + \frac{\sqrt{T}(1-x)}{1-Tx}(\lambda^*\alpha + \omega\beta^*)}.
\label{eq:state_noisychannel}
\end{align}
	
In the case of a loss-only channel, i.e., in absence of additional thermal noise ($x = 0$), Eq. \eqref{eq:state_noisychannel} reduces to
\begin{align}
&\lim_{x\rightarrow 0} U_{\rm{ch}}:\rho_{\rm{s,in}} \rightarrow 
\int \frac{d^2\alpha}{\pi}\frac{d^2\beta}{\pi} C(\alpha,\beta) e^{-\frac{1-T}{2}\argp{|\alpha|^2 + |\beta|^2} + (1-T)\alpha\beta^*}
\int \frac{d^2\lambda}{\pi}\frac{d^2\omega}{\pi} 
e^{-\frac{|\lambda|^2 + |\omega|^2 + T\argp{|\alpha|^2 + |\beta|^2}}{2} + \sqrt{T}(\lambda^*\alpha + \omega\beta^*)} \ket\lambda\bra\omega
\nonumber
\\
&= \int \frac{d^2\alpha}{\pi}\frac{d^2\beta}{\pi} C(\alpha,\beta) e^{-\frac{1-T}{2}\argp{|\alpha|^2 + |\beta|^2} + (1-T)\alpha\beta^*}
\int \frac{d^2\lambda}{\pi}\frac{d^2\omega}{\pi} 
\ket\lambda\bra\lambda\ket{\sqrt{T}\alpha} \bra{\sqrt{T}\beta}\ket\omega\bra\omega
\nonumber
\\
&= \int \frac{d^2\alpha}{\pi}\frac{d^2\beta}{\pi} C(\alpha,\beta) e^{-\frac{1-T}{2}\argp{|\alpha|^2 + |\beta|^2} + (1-T)\alpha\beta^*} \ket{\sqrt{T}\alpha} \bra{\sqrt{T}\beta}.
\end{align}
\end{widetext}

\subsection{Hybrid state after channel transmission} 
\label{subsec:hes_channel}

In our protocol, each party transmits a coherent state through a quantum channel which may have transmission loss as well as thermal noise. In the following, we evaluate all possible terms that may arise when the parties transmit the continuous variable part through the aforementioned channel.
In such a case, Eq. \eqref{eq:state_noisychannel} leads to
\begin{widetext}
    \begin{align}
        U_{\rm{ch}}:\ket\gamma\bra\gamma &\rightarrow \frac{1-x}{1-Tx} \int \frac{d^2\lambda}{\pi}\frac{d^2\omega}{\pi} 
        e^{-\frac{|\lambda|^2 + |\omega|^2}{2} + \frac{x(1-T)}{1-Tx}\lambda^*\omega} \ket\lambda\bra\omega
        \int \frac{d^2\alpha}{\pi}\frac{d^2\beta}{\pi} 
        \delta^2\argp{\beta - \gamma} \delta^2\argp{\alpha - \gamma}
        \nonumber
        \\
        &~~\times e^{-\frac{|\alpha|^2 + |\beta|^2}{2} + \frac{1-T}{1-Tx}\alpha\beta^* + \frac{\sqrt{T}(1-x)}{1-Tx}(\lambda^*\alpha + \omega\beta^*)}
        \nonumber
        \\
        &= \frac{1-x}{1-Tx} \int \frac{d^2\lambda}{\pi}\frac{d^2\omega}{\pi} 
        e^{-\frac{|\lambda|^2 + |\omega|^2}{2} + \frac{x(1-T)}{1-Tx}\lambda^*\omega} \ket\lambda\bra\omega
        \times
        e^{-\frac{T(1-x)}{1-Tx}|\gamma|^2  + \frac{\sqrt{T}(1-x)}{1-Tx}(\lambda^*\gamma + \omega\gamma^*)}
        \nonumber
        \\
        &= e^{-\frac{T(1-x)}{1-Tx}|\gamma|^2} \frac{1-x}{1-Tx} 
        \int \frac{d^2\lambda}{\pi}\frac{d^2\omega}{\pi} 
        e^{-\frac{|\lambda|^2 + |\omega|^2}{2} + \frac{x(1-T)}{1-Tx}\lambda^*\omega + \frac{\sqrt{T}(1-x)}{1-Tx}(\lambda^*\gamma + \omega\gamma^*)} \ket\lambda\bra\omega,
    \end{align}
	
    \begin{align}
        U_{\rm{ch}}:\ket{\gamma}\bra{-\gamma} &\rightarrow \frac{1-x}{1-Tx} 
        \int \frac{d^2\lambda}{\pi}\frac{d^2\omega}{\pi} 
        e^{-\frac{|\lambda|^2 + |\omega|^2}{2} + \frac{x(1-T)}{1-Tx}\lambda^*\omega} \ket\lambda\bra\omega
        \int \frac{d^2\alpha}{\pi}\frac{d^2\beta}{\pi} 
        \delta^2\argp{\beta + \gamma} \delta^2\argp{\alpha - \gamma}
        \nonumber
        \\
        &~~\times e^{-\frac{|\alpha|^2 + |\beta|^2}{2} + \frac{1-T}{1-Tx}\alpha\beta^* + \frac{\sqrt{T}(1-x)}{1-Tx}(\lambda^*\alpha + \omega\beta^*)}
        \nonumber
        \\
        &= \frac{1-x}{1-Tx} \int \frac{d^2\lambda}{\pi}\frac{d^2\omega}{\pi} 
        e^{-\frac{|\lambda|^2 + |\omega|^2}{2} + \frac{x(1-T)}{1-Tx}\lambda^*\omega} \ket\lambda\bra\omega
        \times
        e^{-\frac{T(1-x)}{1-Tx}|\gamma|^2  + \frac{\sqrt{T}(1-x)}{1-Tx}(\lambda^*\gamma - \omega\gamma^*)}
        \nonumber
        \\
        &= e^{-\frac{T(1-x)}{1-Tx}|\gamma|^2} \frac{1-x}{1-Tx} 
        \int \frac{d^2\lambda}{\pi}\frac{d^2\omega}{\pi} 
        e^{-\frac{|\lambda|^2 + |\omega|^2}{2} + \frac{x(1-T)}{1-Tx}\lambda^*\omega + \frac{\sqrt{T}(1-x)}{1-Tx}(\lambda^*\gamma - \omega\gamma^*)} \ket\lambda\bra\omega,
    \end{align}
	
    \begin{align}
        U_{\rm{ch}}:\ket{-\gamma}\bra{\gamma} &\rightarrow \frac{1-x}{1-Tx} 
        \int \frac{d^2\lambda}{\pi}\frac{d^2\omega}{\pi} 
        e^{-\frac{|\lambda|^2 + |\omega|^2}{2} + \frac{x(1-T)}{1-Tx}\lambda^*\omega} \ket\lambda\bra\omega
        \int \frac{d^2\alpha}{\pi}\frac{d^2\beta}{\pi} 
        \delta^2\argp{\beta - \gamma} \delta^2\argp{\alpha + \gamma}
        \nonumber
        \\
        &~~\times e^{-\frac{|\alpha|^2 + |\beta|^2}{2} + \frac{1-T}{1-Tx}\alpha\beta^* + \frac{\sqrt{T}(1-x)}{1-Tx}(\lambda^*\alpha + \omega\beta^*)}
        \nonumber
        \\
        &= \frac{1-x}{1-Tx} \int \frac{d^2\lambda}{\pi}\frac{d^2\omega}{\pi} 
        e^{-\frac{|\lambda|^2 + |\omega|^2}{2} + \frac{x(1-T)}{1-Tx}\lambda^*\omega} \ket\lambda\bra\omega
        \times
        e^{-\frac{T(1-x)}{1-Tx}|\gamma|^2  + \frac{\sqrt{T}(1-x)}{1-Tx}(-\lambda^*\gamma + \omega\gamma^*)}
        \nonumber
        \\
        &= e^{-\frac{T(1-x)}{1-Tx}|\gamma|^2} \frac{1-x}{1-Tx} 
        \int \frac{d^2\lambda}{\pi}\frac{d^2\omega}{\pi} 
        e^{-\frac{|\lambda|^2 + |\omega|^2}{2} + \frac{x(1-T)}{1-Tx}\lambda^*\omega - \frac{\sqrt{T}(1-x)}{1-Tx}(\lambda^*\gamma - \omega\gamma^*)} \ket\lambda\bra\omega,
    \end{align}
	
    \begin{align}
        U_{\rm{ch}}:\ket{-\gamma}\bra{-\gamma} &\rightarrow \frac{1-x}{1-Tx} 
        \int \frac{d^2\lambda}{\pi}\frac{d^2\omega}{\pi} 
        e^{-\frac{|\lambda|^2 + |\omega|^2}{2} + \frac{x(1-T)}{1-Tx}\lambda^*\omega} \ket\lambda\bra\omega
        \int \frac{d^2\alpha}{\pi}\frac{d^2\beta}{\pi} 
        \delta^2\argp{\beta + \gamma} \delta^2\argp{\alpha + \gamma}
        \nonumber
        \\
        &~~\times e^{-\frac{|\alpha|^2 + |\beta|^2}{2} + \frac{1-T}{1-Tx}\alpha\beta^* + \frac{\sqrt{T}(1-x)}{1-Tx}(\lambda^*\alpha + \omega\beta^*)}
        \nonumber
        \\
        &= \frac{1-x}{1-Tx} \int \frac{d^2\lambda}{\pi}\frac{d^2\omega}{\pi} 
        e^{-\frac{|\lambda|^2 + |\omega|^2}{2} + \frac{x(1-T)}{1-Tx}\lambda^*\omega} \ket\lambda\bra\omega
        \times
        e^{-\frac{T(1-x)}{1-Tx}|\gamma|^2  - \frac{\sqrt{T}(1-x)}{1-Tx}(\lambda^*\gamma + \omega\gamma^*)}
        \nonumber
        \\
        &= e^{-\frac{T(1-x)}{1-Tx}|\gamma|^2} \frac{1-x}{1-Tx} 
        \int \frac{d^2\lambda}{\pi}\frac{d^2\omega}{\pi} 
        e^{-\frac{|\lambda|^2 + |\omega|^2}{2} + \frac{x(1-T)}{1-Tx}\lambda^*\omega - \frac{\sqrt{T}(1-x)}{1-Tx}(\lambda^*\gamma + \omega\gamma^*)} \ket\lambda\bra\omega.
    \end{align}
\end{widetext}  
	
A hybrid-entangled (HE) state \eqref{eq:hes} defined as
\begin{equation}
\ket{\psi}_{\rm{he}} = \frac{1}{\sqrt{2}} \argp{\ket{0,\alpha} + \ket{1,-\alpha}},
\end{equation}
for which the multiphoton coherent state part is transmitted through a general (both lossy and noisy) channel. 
Using the results obtained above it can be seen that, after the transmission, the final state is
\begin{widetext}
    \begin{align}
        &\rho^{\rm{ch,he}}(T,x) = U_{\rm{ch}}: \rho_{\rm{he}} = U_{\rm{ch}}: \frac{1}{2} \argp{\ket 0\bra 0 \otimes \ket\alpha\bra\alpha + \ket 0\bra 1 \otimes \ket{\alpha}\bra{-\alpha} + \ket 1\bra 0 \ket{-\alpha}\bra{\alpha} + \ket 1\bra 1 \otimes \ket{-\alpha}\bra{-\alpha}}
        \nonumber
        \\
        &= \frac{e^{-\frac{T(1-x)}{1-Tx}|\alpha|^2}}{2} \frac{1-x}{1-Tx} 
        \int \frac{d^2\lambda}{\pi}\frac{d^2\omega}{\pi} 
        e^{-\frac{|\lambda|^2 + |\omega|^2}{2} + \frac{x(1-T)}{1-Tx}\lambda^*\omega} 
        \ket\lambda\bra\omega
        \nonumber 
        \\
        &~~\otimes 
        \args{e^{\frac{\sqrt{T}(1-x)}{1-Tx}(\lambda^*\alpha + \omega\alpha^*)}\ket 0\bra 0 + 
	e^{\frac{\sqrt{T}(1-x)}{1-Tx}(\lambda^*\alpha - \omega\alpha^*)}\ket 0\bra 1 + 
	e^{-\frac{\sqrt{T}(1-x)}{1-Tx}(\lambda^*\alpha - \omega\alpha^*)}\ket 1\bra 0 +
	e^{-\frac{\sqrt{T}(1-x)}{1-Tx}(\lambda^*\alpha + \omega\alpha^*)}\ket 1\bra 1}.	 
        \label{eq:hes_channel}
    \end{align}
\end{widetext}  

\subsection{HE-state at Charlie's end after transmission through loss-only channel and a general channel}
\label{subsec:hes_charlie_channel}

\begin{figure}
    \includegraphics[scale=0.9]{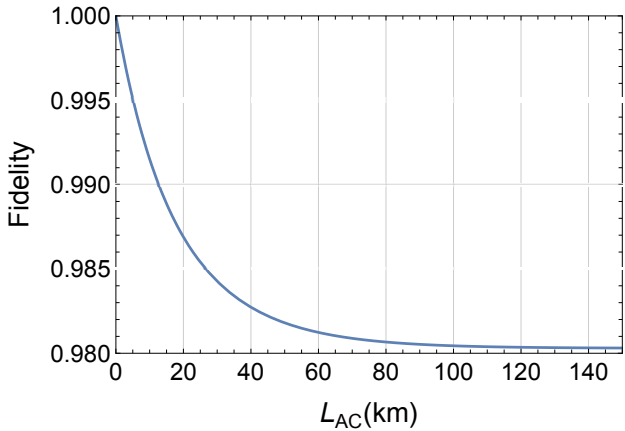}
    \caption{Dependence of fidelity between the states obtained from initial HE state after transmission through loss-only and both loss$+$noise channels.}
    \label{fig:hes_channel_fid_lossnoise}
\end{figure}

Let us consider that Alice and Bob prepare their individual HE states $\ket\psi_{a_1a_2}$ and $\ket\psi_{b_1b_2}$ as
\begin{subequations}
    \begin{align}
        \ket\psi_{a_1a_2} &= \frac{1}{\sqrt{2}} \argp{\ket{0,\alpha}_{a_1a_2} + \ket{1,-\alpha}_{a_1a_2}} 
        \\
        \ket\psi_{b_1b_2} &= \frac{1}{\sqrt{2}} \argp{\ket{0,\alpha}_{b_1b_2} + \ket{1,-\alpha}_{b_1b_2}},
    \end{align}
\end{subequations}
where $\ket{0,\alpha}_{a_1a_2}=\ket 0_{a_1}\ket\alpha_{a_2}$. 
Using Eq.~\eqref{eq:hes_channel} one can show that total $4$-mode state at the input of Charlie after passing through a general channel is given by
\begin{widetext}
    \begin{align}
        &\rho_{\rm{in,tot}}(T,x) = \rho_{a_1a_2}^{\rm{ch}} \otimes \rho_{b_1b_2}^{\rm{ch}}
        \nonumber
        \\
        &= \frac{e^{-\frac{2T(1-x)}{1-Tx}|\alpha|^2}}{4}  \argp{\frac{1-x}{1-Tx}}^2
        \int \frac{d^2\lambda}{\pi}\frac{d^2\omega}{\pi} \frac{d^2\chi}{\pi}\frac{d^2\xi}{\pi}
        e^{-\frac{|\lambda|^2 + |\omega|^2 + |\chi|^2 + |\xi|^2}{2} + \frac{x(1-T)}{1-Tx}\argp{\lambda^*\omega + \chi^*\xi}} 
        \ket\lambda_{a_2}\bra\omega \otimes \ket\chi_{b_2}\bra\xi
        \nonumber 
        \\
        &~~\otimes 
        \args{e^{\frac{\sqrt{T}(1-x)}{1-Tx}(\lambda^*\alpha + \omega\alpha^*)}\ket 0_{a_1}\bra 0 + 
        e^{\frac{\sqrt{T}(1-x)}{1-Tx}(\lambda^*\alpha - \omega\alpha^*)}\ket 0_{a_1}\bra 1 + 
        e^{-\frac{\sqrt{T}(1-x)}{1-Tx}(\lambda^*\alpha - \omega\alpha^*)}\ket 1_{a_1}\bra 0 +
        e^{-\frac{\sqrt{T}(1-x)}{1-Tx}(\lambda^*\alpha + \omega\alpha^*)}\ket 1_{a_1}\bra 1}
        \nonumber 
        \\
        &~~\otimes 
        \args{e^{\frac{\sqrt{T}(1-x)}{1-Tx}(\chi^*\alpha + \xi\alpha^*)}\ket 0_{b_1}\bra 0 + 
	e^{\frac{\sqrt{T}(1-x)}{1-Tx}(\chi^*\alpha - \xi\alpha^*)}\ket 0_{b_1}\bra 1 + 
	e^{-\frac{\sqrt{T}(1-x)}{1-Tx}(\chi^*\alpha - \xi\alpha^*)}\ket 1_{b_1}\bra 0 +
	e^{-\frac{\sqrt{T}(1-x)}{1-Tx}(\chi^*\alpha + \xi\alpha^*)}\ket 1_{b_1}\bra 1}
        \nonumber
        \\
        &= \frac{e^{-\frac{2T(1-x)}{1-Tx}|\alpha|^2}}{4}  \argp{\frac{1-x}{1-Tx}}^2
        \int \frac{d^2\lambda}{\pi}\frac{d^2\omega}{\pi} \frac{d^2\chi}{\pi}\frac{d^2\xi}{\pi}
        e^{-\frac{|\lambda|^2 + |\omega|^2 + |\chi|^2 + |\xi|^2}{2} + \frac{x(1-T)}{1-Tx}\argp{\lambda^*\omega + \chi^*\xi}} 
        \ket{\lambda,\chi}_{a_2b_2}\bra{\omega,\xi}
        \nonumber 
        \\
        &~~\otimes 
        \left[ \left(
        e^{\frac{\sqrt{T}(1-x)}{1-Tx}\args{(\lambda^* + \chi^*)\alpha + (\omega + \xi)\alpha^*}} \ket{0,0}_{a_1b_1}\bra{0,0}
        +
        e^{\frac{\sqrt{T}(1-x)}{1-Tx}\args{(\lambda^* + \chi^*)\alpha + (\omega - \xi)\alpha^*}} \ket{0,0}_{a_1b_1}\bra{0,1}
        \right.
        \right.
        \nonumber 
        \\
        &~~\left. +
        e^{\frac{\sqrt{T}(1-x)}{1-Tx}\args{(\lambda^* - \chi^*)\alpha + (\omega + \xi)\alpha^*}} \ket{0,1}_{a_1b_1}\bra{0,0}
        +
        e^{\frac{\sqrt{T}(1-x)}{1-Tx}\args{(\lambda^* - \chi^*)\alpha + (\omega - \xi)\alpha^*}} \ket{0,1}_{a_1b_1}\bra{0,1}
        \right)
        \nonumber 
        \\
        &~~ + \left(
        e^{\frac{\sqrt{T}(1-x)}{1-Tx}\args{(\lambda^* + \chi^*)\alpha - (\omega - \xi)\alpha^*}} \ket{0,0}_{a_1b_1}\bra{1,0}
        +
        e^{\frac{\sqrt{T}(1-x)}{1-Tx}\args{(\lambda^* + \chi^*)\alpha - (\omega + \xi)\alpha^*}} \ket{0,0}_{a_1b_1}\bra{1,1}
        \right.
        \nonumber 
        \\
        &~~\left. +
        e^{\frac{\sqrt{T}(1-x)}{1-Tx}\args{(\lambda^* - \chi^*)\alpha - (\omega - \xi)\alpha^*}} \ket{0,1}_{a_1b_1}\bra{1,0}
        +
        e^{\frac{\sqrt{T}(1-x)}{1-Tx}\args{(\lambda^* - \chi^*)\alpha - (\omega + \xi)\alpha^*}} \ket{0,1}_{a_1b_1}\bra{1,1}
        \right)
        \nonumber 
        \\
        &~~ +\left(
        e^{-\frac{\sqrt{T}(1-x)}{1-Tx}\args{(\lambda^* - \chi^*)\alpha - (\omega + \xi)\alpha^*}} \ket{1,0}_{a_1b_1}\bra{0,0}
        +
        e^{-\frac{\sqrt{T}(1-x)}{1-Tx}\args{(\lambda^* - \chi^*)\alpha - (\omega - \xi)\alpha^*}} \ket{1,0}_{a_1b_1}\bra{0,1}
        \right.
        \nonumber 
        \\
        &~~\left. +
        e^{-\frac{\sqrt{T}(1-x)}{1-Tx}\args{(\lambda^* + \chi^*)\alpha - (\omega + \xi)\alpha^*}} \ket{1,1}_{a_1b_1}\bra{0,0}
        +
        e^{-\frac{\sqrt{T}(1-x)}{1-Tx}\args{(\lambda^* + \chi^*)\alpha - (\omega - \xi)\alpha^*}} \ket{1,1}_{a_1b_1}\bra{0,1}
        \right)
        \nonumber 
        \\
        &~~ +\left(
        e^{-\frac{\sqrt{T}(1-x)}{1-Tx}\args{(\lambda^* - \chi^*)\alpha + (\omega - \xi)\alpha^*}} \ket{1,0}_{a_1b_1}\bra{1,0}
        +
        e^{-\frac{\sqrt{T}(1-x)}{1-Tx}\args{(\lambda^* - \chi^*)\alpha + (\omega + \xi)\alpha^*}} \ket{1,0}_{a_1b_1}\bra{1,1}
        \right.
        \nonumber 
        \\
        &~~\left.\left. +
        e^{-\frac{\sqrt{T}(1-x)}{1-Tx}\args{(\lambda^* + \chi^*)\alpha + (\omega - \xi)\alpha^*}} \ket{1,1}_{a_1b_1}\bra{1,0}
        +
        e^{-\frac{\sqrt{T}(1-x)}{1-Tx}\args{(\lambda^* + \chi^*)\alpha + (\omega + \xi)\alpha^*}} \ket{1,1}_{a_1b_1}\bra{1,1}
        \right)\right].
        \label{eq:hes_Charlie_gen}
    \end{align}

    It should be noted, in absence of the additional thermal noise ($x\rightarrow 0$) the transmission channel simply becomes simply a loss-only channel and the state at the input of Charlie (after transmission) can be written as
    \begin{align}
        &\rho_{\rm{in,tot}}(T) = \lim_{x\rightarrow 0} \rho_{\rm{in,tot}}(T,x) 
        \nonumber
        \\
        &= \frac{e^{-2T|\alpha|^2}}{4}
        \int \frac{d^2\lambda}{\pi}\frac{d^2\omega}{\pi} \frac{d^2\chi}{\pi}\frac{d^2\xi}{\pi}
        e^{-\frac{|\lambda|^2 + |\omega|^2 + |\chi|^2 + |\xi|^2}{2}} 
        \ket{\lambda,\chi}_{a_2b_2}\bra{\omega,\xi}
        \nonumber 
        \\
        &~~\otimes 
        \left[ \left(
        e^{\sqrt{T}\args{(\lambda^* + \chi^*)\alpha + (\omega + \xi)\alpha^*}} \ket{0,0}_{a_1b_1}\bra{0,0}
        +
        e^{\sqrt{T}\args{(\lambda^* + \chi^*)\alpha + (\omega - \xi)\alpha^*}} \ket{0,0}_{a_1b_1}\bra{0,1}
        +
        e^{\sqrt{T}\args{(\lambda^* - \chi^*)\alpha + (\omega + \xi)\alpha^*}} \ket{0,1}_{a_1b_1}\bra{0,0}
        \right.\right.
        \nonumber 
        \\
        &~~\left. +
        e^{\sqrt{T}\args{(\lambda^* - \chi^*)\alpha + (\omega - \xi)\alpha^*}} \ket{0,1}_{a_1b_1}\bra{0,1}
        \right)
        +
        \left(
        e^{\sqrt{T}\args{(\lambda^* + \chi^*)\alpha - (\omega - \xi)\alpha^*}} \ket{0,0}_{a_1b_1}\bra{1,0}
        +
        e^{\sqrt{T}\args{(\lambda^* + \chi^*)\alpha - (\omega + \xi)\alpha^*}} \ket{0,0}_{a_1b_1}\bra{1,1}
        \right.
        \nonumber 
        \\
        &~~\left. +
        e^{\sqrt{T}\args{(\lambda^* - \chi^*)\alpha - (\omega - \xi)\alpha^*}} \ket{0,1}_{a_1b_1}\bra{1,0}
        +
        e^{\sqrt{T}\args{(\lambda^* - \chi^*)\alpha - (\omega + \xi)\alpha^*}} \ket{0,1}_{a_1b_1}\bra{1,1}
        \right)
        \nonumber 
        \\
        &~~ +\left(
        e^{-\sqrt{T}\args{(\lambda^* - \chi^*)\alpha - (\omega + \xi)\alpha^*}} \ket{1,0}_{a_1b_1}\bra{0,0}
        +
        e^{-\sqrt{T}\args{(\lambda^* - \chi^*)\alpha - (\omega - \xi)\alpha^*}} \ket{1,0}_{a_1b_1}\bra{0,1}
        +
        e^{-\sqrt{T}\args{(\lambda^* + \chi^*)\alpha - (\omega + \xi)\alpha^*}} \ket{1,1}_{a_1b_1}\bra{0,0}
        \right.
        \nonumber 
        \\
        &~~\left. +
        e^{-\sqrt{T}\args{(\lambda^* + \chi^*)\alpha - (\omega - \xi)\alpha^*}} \ket{1,1}_{a_1b_1}\bra{0,1}
        \right)
        +
        \left(
        e^{-\sqrt{T}\args{(\lambda^* - \chi^*)\alpha + (\omega - \xi)\alpha^*}} \ket{1,0}_{a_1b_1}\bra{1,0}
        +
        e^{-\sqrt{T}\args{(\lambda^* - \chi^*)\alpha + (\omega + \xi)\alpha^*}} \ket{1,0}_{a_1b_1}\bra{1,1}
        \right.
        \nonumber 
        \\
        &~~\left.\left. +
        e^{-\sqrt{T}\args{(\lambda^* + \chi^*)\alpha + (\omega - \xi)\alpha^*}} \ket{1,1}_{a_1b_1}\bra{1,0}
        +
        e^{-\sqrt{T}\args{(\lambda^* + \chi^*)\alpha + (\omega + \xi)\alpha^*}} \ket{1,1}_{a_1b_1}\bra{1,1}
        \right)\right].
        \label{eq:hes_Charlie_loss}
    \end{align}
\end{widetext}
	
\subsection{Fidelity between HE-states at Charlie's end after transmission through loss-only channel and a general channel}
\label{subsec:hes_fid_charlie_channel}

In this subsection we evaluate how different the state given in Eq.~\eqref{eq:hes_Charlie_gen} is from the one given in Eq.~\eqref{eq:hes_Charlie_loss} for a given loss and thermal noise.
In order to estimate it we evaluate the fidelity between the states given by Eq.~\eqref{eq:hes_Charlie_gen} and Eq.~\eqref{eq:hes_Charlie_loss} as
\begin{widetext}
    \begin{align}
&F = \rm{tr}\args{\rho_{\rm{in,tot}}(T) \rho_{\rm{in,tot}}(T,x)}
\nonumber
\\
&= \frac{e^{-2T\frac{2-x(1+T)}{1-Tx}|\alpha|^2}}{16} \argp{\frac{1-x}{1-Tx}}^2 
\int \frac{d^2\Lambda_i}{\pi^8}
e^{-\argp{|\lambda_1|^2 + |\omega_1|^2 + |\chi_1|^2 + |\xi_1|^2 + |\lambda_2|^2 + |\omega_2|^2 + |\chi_2|^2 + |\xi_2|^2} + \argp{\omega_2^*\lambda_1 + \xi_2^*\chi_1 + \omega_1^*\lambda_2 + \xi_1^*\chi_2}}
\nonumber
\\
&~~\times	\left[\left(
e^{\frac{\sqrt{T}(1-x)}{1-Tx}\args{(\lambda_1^* + \chi_1^*)\alpha + (\omega_1 + \xi)\alpha_1^*} + \sqrt{T}\args{(\lambda_2^* + \chi_2^*)\alpha + (\omega_2 + \xi_2)\alpha^*}} 
+
e^{\frac{\sqrt{T}(1-x)}{1-Tx}\args{(\lambda_1^* + \chi_1^*)\alpha + (\omega_1 - \xi)\alpha_1^*} + \sqrt{T}\args{(\lambda_2^* - \chi_2^*)\alpha + (\omega_2 + \xi_2)\alpha^*}}
\right.\right.
\nonumber 
\\
&~~	+\left.	
e^{\frac{\sqrt{T}(1-x)}{1-Tx}\args{(\lambda_1^* - \chi_1^*)\alpha + (\omega_1 + \xi)\alpha_1^*} + \sqrt{T}\args{(\lambda_2^* + \chi_2^*)\alpha + (\omega_2 - \xi_2)\alpha^*}}
+
e^{\frac{\sqrt{T}(1-x)}{1-Tx}\args{(\lambda_1^* - \chi_1^*)\alpha + (\omega_1 - \xi)\alpha_1^*} + \sqrt{T}\args{(\lambda_2^* - \chi_2^*)\alpha + (\omega_2 - \xi_2)\alpha^*}}
\right)
\nonumber 
\\
&~~ + \left(
e^{\frac{\sqrt{T}(1-x)}{1-Tx}\args{(\lambda_1^* + \chi_1^*)\alpha - (\omega_1 - \xi_1)\alpha^*} - \sqrt{T}\args{(\lambda_2^* - \chi_2^*)\alpha - (\omega_2 + \xi_2)\alpha^*}} 
+
e^{\frac{\sqrt{T}(1-x)}{1-Tx}\args{(\lambda_1^* + \chi_1^*)\alpha - (\omega_1 + \xi_1)\alpha^*} - \sqrt{T}\args{(\lambda_2^* + \chi_2^*)\alpha - (\omega_2 + \xi_2)\alpha^*}} 
\right.
\nonumber 
\\
&~~\left. +
e^{\frac{\sqrt{T}(1-x)}{1-Tx}\args{(\lambda_1^* - \chi_1^*)\alpha - (\omega_1 - \xi_1)\alpha^*} - \sqrt{T}\args{(\lambda_2^* - \chi_2^*)\alpha - (\omega_2 - \xi_2)\alpha^*}} 
+
e^{\frac{\sqrt{T}(1-x)}{1-Tx}\args{(\lambda_1^* - \chi_1^*)\alpha - (\omega_1 + \xi_1)\alpha^*} - \sqrt{T}\args{(\lambda_2^* + \chi_2^*)\alpha - (\omega_2 - \xi_2)\alpha^*}} 
\right)
\nonumber 
\\
&~~ +\left(
e^{-\frac{\sqrt{T}(1-x)}{1-Tx}\args{(\lambda_1^* - \chi_1^*)\alpha - (\omega_1 + \xi_1)\alpha^*} + \sqrt{T}\args{(\lambda_2^* + \chi_2^*)\alpha - (\omega_2 - \xi_2)\alpha^*}} 
+
e^{-\frac{\sqrt{T}(1-x)}{1-Tx}\args{(\lambda_1^* - \chi_1^*)\alpha - (\omega_1 - \xi_1)\alpha^*} + \sqrt{T}\args{(\lambda_2^* - \chi_2^*)\alpha - (\omega_2 - \xi_2)\alpha^*}} 
\right.
\nonumber 
\\
&~~\left. +
e^{-\frac{\sqrt{T}(1-x)}{1-Tx}\args{(\lambda_1^* + \chi_1^*)\alpha - (\omega_1 + \xi_1)\alpha^*} + \sqrt{T}\args{(\lambda_2^* + \chi_2^*)\alpha - (\omega_2 + \xi_2)\alpha^*}} 
+
e^{-\frac{\sqrt{T}(1-x)}{1-Tx}\args{(\lambda_1^* + \chi_1^*)\alpha - (\omega_1 - \xi_1)\alpha^*} + \sqrt{T}\args{(\lambda_2^* - \chi_2^*)\alpha - (\omega_2 + \xi_2)\alpha^*}} 
\right)
\nonumber 
\\
&~~ +\left(
e^{-\frac{\sqrt{T}(1-x)}{1-Tx}\args{(\lambda_1^* - \chi_1^*)\alpha + (\omega_1 - \xi_1)\alpha^*} - \sqrt{T}\args{(\lambda_2^* - \chi_2^*)\alpha + (\omega_2 - \xi_2)\alpha^*}} 
+
e^{-\frac{\sqrt{T}(1-x)}{1-Tx}\args{(\lambda_1^* - \chi_1^*)\alpha + (\omega_1 + \xi_1)\alpha^*} - \sqrt{T}\args{(\lambda_2^* + \chi_2^*)\alpha + (\omega_2 - \xi_2)\alpha^*}} 
\right.
\nonumber 
\\
&~~\left.\left. +
e^{-\frac{\sqrt{T}(1-x)}{1-Tx}\args{(\lambda_1^* + \chi_1^*)\alpha + (\omega_1 - \xi_1)\alpha^*} - \sqrt{T}\args{(\lambda_2^* - \chi_2^*)\alpha + (\omega_2 + \xi_2)\alpha^*}} 
+
e^{-\frac{\sqrt{T}(1-x)}{1-Tx}\args{(\lambda_1^* + \chi_1^*)\alpha + (\omega_1 + \xi_1)\alpha^*} - \sqrt{T}\args{(\lambda_2^* + \chi_2^*)\alpha + (\omega_2 + \xi_2)\alpha^*}} 
\right)\right]
\label{eq:hes_channel_fid_preli},
\end{align}
where $d^2\Lambda_i = d^2\lambda_i d^2\omega_i d^2\chi_i d^2\xi_i$ ($i=1,2$).
	
Let us now consider a generic integral as
\begin{align}
&I_1 = \frac{e^{-2T\frac{2-x(1+T)}{1-Tx}|\alpha|^2}}{16} \argp{\frac{1-x}{1-Tx}}^2 
\int \frac{d^2\Lambda_i}{\pi^8}
e^{-\argp{|\lambda_1|^2 + |\omega_1|^2 + |\chi_1|^2 + |\xi_1|^2 + |\lambda_2|^2 + |\omega_2|^2 + |\chi_2|^2 + |\xi_2|^2} + \argp{\omega_2^*\lambda_1 + \xi_2^*\chi_1 + \omega_1^*\lambda_2 + \xi_1^*\chi_2}}
\nonumber 
\\
&~~\times 
e^{\args{(A_1\lambda_1^* + B_1\chi_1^*)\alpha + (C_1\omega_1 + D_1\xi_1)\alpha_1^*} + \args{(A_2\lambda_2^* + B_2\chi_2^*)\alpha + (C_2\omega_2 + D_2\xi_2)\alpha^*}} 
\nonumber
\\
&= 	\frac{e^{-2T\frac{2-x(1+T)}{1-Tx}|\alpha|^2}}{16} \argp{\frac{1-x}{1-Tx}}^2 
\int \frac{d^2\lambda_1}{\pi}\frac{d^2\omega_1}{\pi} \frac{d^2\chi_1}{\pi}\frac{d^2\xi_1}{\pi}
e^{-\argp{|\lambda_1|^2 + |\omega_1|^2 + |\chi_1|^2 + |\xi_1|^2} + (A_1\lambda_1^* + B_1\chi_1^*)\alpha + (C_1\omega_1 + D_1\xi_1)\alpha_1^*}
\nonumber 
\\
&~~\times
\int \frac{d^2\lambda_2}{\pi}\frac{d^2\omega_2}{\pi} \frac{d^2\chi_2}{\pi}\frac{d^2\xi_2}{\pi}
e^{-\argp{|\lambda_2|^2 + |\omega_2|^2 + |\chi_2|^2 + |\xi_2|^2} + \argp{\omega_2^*\lambda_1 + \xi_2^*\chi_1 + \omega_1^*\lambda_2 + \xi_1^*\chi_2} + (A_2\lambda_2^* + B_2\chi_2^*)\alpha + (C_2\omega_2 + D_2\xi_2)\alpha^*}
\nonumber
\\
&= 	\frac{e^{-2T\frac{2-x(1+T)}{1-Tx}|\alpha|^2}}{16} \argp{\frac{1-x}{1-Tx}}^2 
\int \frac{d^2\lambda_1}{\pi}\frac{d^2\omega_1}{\pi} \frac{d^2\chi_1}{\pi}\frac{d^2\xi_1}{\pi}
e^{-\argp{|\lambda_1|^2 + |\omega_1|^2 + |\chi_1|^2 + |\xi_1|^2} + (A_1\lambda_1^* + B_1\chi_1^*)\alpha + (C_1\omega_1 + D_1\xi_1)\alpha_1^*}
\nonumber 
\\
&~~\times
e^{(A_2\omega_1^* + B_2\xi_1^*)\alpha + (C_2\lambda_1 + D_2\chi_1)\alpha^*}
\nonumber
\\
&= \frac{e^{-2T\frac{2-x(1+T)}{1-Tx}|\alpha|^2}}{16} \argp{\frac{1-x}{1-Tx}}^2 e^{\args{\argp{A_1C_2 + A_2C_1} + \argp{B_1D_2 + B_2D_1}}|\alpha|^2}
\label{eq:hes_channel_fid_genint}.
\end{align}
\end{widetext}

Using the result of the generic integral \eqref{eq:hes_channel_fid_genint}, from \eqref{eq:hes_channel_fid_preli} we get
\begin{align}
    F &= \frac{e^{-2T\frac{2-x(1+T)}{1-Tx}|\alpha|^2}}{16} \argp{\frac{1-x}{1-Tx}}^2 
    \times
    16 e^{4T\frac{(1-x)}{1-Tx}} 
    \nonumber
    \\
    &= e^{-\frac{2Tx(1-T)}{1-Tx}|\alpha|^2} \argp{\frac{1-x}{1-Tx}}^2.
    \label{eq:hes_channel_fid_final}
\end{align}

This analysis is important because a fully general calculation, in which we consider the quantum channels to be characterized by transmission loss and thermal noise is far more involved and lengthy to perform than if we only consider the quantum channels to be characterized by transmission loss only. 
This can be seen from the form of the $4$-mode states at Charlie's input before the entanglement-swapping operation.
In view of the result represented in Fig. \mbox{\ref{fig:hes_channel_fid_lossnoise}}, we believe that a consideration of such a general channel may not yield any significantly different result from a loss-only channel at the cost of a very difficult, lengthy and complicated calculation.
It may be noted that in the case of a loss-only channel the ancilla thermal state is replaced by a vacuum state. 
This simplifies the calculation greatly as now we can proceed with a pure state approach in which the total state is a pure state. 
In this case, it is possible to take a partial trace over the ancilla after Charlie's operations. 
This simplifies the overall calculation. However, such a simplification is not possible when we consider an ancilla in the thermal state for which the overall state is mixed. 
This increases the number of terms to be calculated by four times in comparison with the pure state approach.


\begin{thebibliography}{70}
\bibitem [{ono()}]{onoff_detect}%
	\BibitemOpen
	\bibinfo {title} {While a photon-number-resolving-detector ({PNRD}) measures
		the number of photons detected, an on-off detector only indicates whether a
		photon is detected or not. {A}s a consequence, an on-off detector is much
		less demanding than a {PNRD}. {R}ecent advances in technology has led to
		development of single-photon detectors with efficiencies as high as
		$0.84$--$0.91$~\cite{PSL12, WWL21, MMC21} with the use of
		high-contrast-grating in nanophotonics settings. {T}his, in turn, implies
		that the efficiency of an on-off detector can be safely approximated up to
		$0.80$ (if not more).}\BibitemShut {Stop}%
\end{thebibliography}
\end{document}